\documentclass[]{emulateapj}
\usepackage[first,bottomafter,light,dvips]{draftcopy}
\usepackage{natbib}
\slugcomment{For ApJ. Accepted January 3, 2005}

\shorttitle{VLBA observations of WR\thinspace140}
\shortauthors{Dougherty et al.}

\def\deg{\ifmmode^\circ\else$^\circ$\fi}
\def\kms{\ifmmode\,km\/s\else$\,{\rm km\/s}$\fi\ }

\def\GHz{\,GHz}
\def\WR140{WR\thinspace140}

\begin{document}
\title{High resolution radio observations of the colliding-wind binary WR\thinspace140}

\author{S.M. Dougherty}
\affil{National Research Council, Herzberg Institute for Astrophysics, Dominion Radio Astrophysical Observatory,\\ P.O. Box 248, Penticton, BC, Canada}
\email{sean.dougherty@nrc.ca}
\author{A.J. Beasley}
\affil{Joint ALMA Office, El Golf 40, 18th floor, Las Condes  7550107, Santiago, Chile}
\author{M.J. Claussen}
\affil{National Radio Astronomy Observatory,
1003 Lopezville Rd., Socorro NM 87801, USA}
\author{B.A. Zauderer}
\affil{Department of Astronomy, University of Maryland,
College Park, MD 20742, USA}
\and
\author{N.J. Bolingbroke\altaffilmark{1}}
\affil{Department of Physics and Astronomy, University of Victoria, 3800 Finnerty Rd, Victoria, BC, Canada}
%% alternative affiliations
\altaffiltext{1}{National Research Council, D.R.A.O}
\begin{abstract}
Milli-arcsecond resolution Very Long Baseline Array (VLBA)
observations of the archetype WR+O star colliding-wind binary (CWB)
system WR\thinspace140 are presented for 23 epochs between orbital
phases 0.74 and 0.97.  At 8.4 GHz, the emission in the wind-collision
region (WCR) is clearly resolved as a bow-shaped arc that rotates as
the orbit progresses. We interpret this rotation as due to the O star
moving from SE to approximately E of the WR star, which leads to
solutions for the orbit inclination of $122\deg\pm5\deg$, the
longitude of the ascending node of $353\deg\pm3\deg$, and an orbit
semi-major axis of $9.0\pm0.5$~mas. The distance to WR\thinspace140 is
determined to be $1.85\pm0.16$~kpc, which requires the O star to be a
supergiant. The inclination implies the mass of the WR and O star to be
$20\pm4$~M$_\odot$ and $54\pm10$~M$_\odot$ respectively. We determine
a wind-momentum ratio of 0.22, with an expected half-opening angle for
the WCR of $63\deg$, consistent with $65\deg\pm10\deg$ derived from
the VLBA observations.  Total flux measurements from Very Large Array
(VLA) observations show the radio emission from WR\thinspace140 is
very closely the same from one orbit to the next, pointing strongly
toward emission, absorption and cooling mechanism(s) that are
controlled largely by the orbital motion. The synchrotron spectra
evolve dramatically through the orbital phases observed, exhibiting
both optically thin and optically thick emission. We discuss a number
of absorption and cooling mechanisms that may determine the evolution
of the synchrotron spectrum with orbital phase.
\end{abstract}


\section{Introduction}
Wolf-Rayet (WR) stars have dense stellar winds from which free-free
continuum emission is observed from near-IR to radio wavelengths. A
significant proportion ($\sim$ a quarter) of these objects also
exhibit non-thermal radio emission \citep{Leitherer:1995,
Leitherer:1997, Chapman:1999, Dougherty:2000b, Cappa:2004}, which
requires the presence of a population of relativistic electrons, in
addition to a magnetic field. The acceleration of particles in these
objects is generally attributed to first-order Fermi acceleration in
shocks within the stellar winds, arising either from wind
instabilities \citep{Chen:1994} or, in the case of binary systems,
collisions of the stellar winds of the two stars \citep{Eichler:1993}.
The colliding-wind binary (CWB) model is strongly supported by
high-resolution radio, optical and infrared imaging of the very wide
WR+OB binaries WR\thinspace146 and WR\thinspace147, where the
non-thermal emission is shown to arise between the WR and OB stars at
the point of ram-pressure balance between the two stellar winds
\citep{Dougherty:1996, Williams:1997, Niemela:1998, Dougherty:2000a}.

The archetype of CWB systems is the 7.9-year period WR+O binary system
WR\thinspace140 (HD\thinspace193793). This binary comprises a
WC7 star and an O4-5 star in a highly elliptical orbit
($e\approx0.88$), where the stellar separation varies between
$\sim2$~AU at periastron to $\sim30$~AU at apastron. This highly
eccentric orbit clearly modulates the dramatic variations in the
emission from the system, observed from X-ray to radio wavelengths
\citep[see][]{Williams:1990}. The X-ray luminosity rises steadily
toward periastron passage, while spectra taken just after periastron
clearly show extra absorption that is visible until about phase 0.1
\citep{Pollock:2005}.  The near-IR emission is characterized by a
sudden peak in emission shortly after periastron, followed by a slower
decline. This behavior is attributed to dust formation in the WCR at
some point shortly after periastron, and its subsequent cooling
\citep{Williams:2002}.

\begin{figure}[t]
\includegraphics[angle=0,scale=0.7]{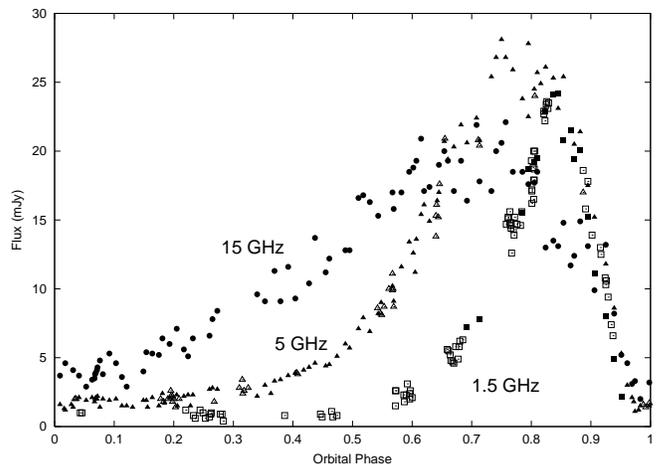}
\caption[]{Radio emission from WR\thinspace140 at 15 (circles), 5
(triangles), and 1.5 GHz (squares) as measured with the VLA (solid)
and WSRT (open) during the orbital cycle between 1984.3 and
1992.2. Data taken from \citet{Williams:1990, Williams:1994} and
\citet{White:1995}.\label{fig:light_curve}}
\end{figure}

Perhaps the most dramatic variations are observed at radio wavelengths
(Fig.~\ref{fig:light_curve}), where there is a slow rise from a low,
apparently thermal state close to periastron of a few mJy, to a
frequency-dependent peak in emission of several 10s of mJy between
orbital phase ($\phi$) 0.65 to 0.85, before a precipitous decline
before periastron \citep{Williams:1990, White:1995}. The radio
variations have been widely attributed to an underlying synchrotron
source viewed through the changing free-free opacity of the extended
stellar wind envelopes of the binary system along the line-of-sight to
the WCR as the orbit progresses \citep{Williams:1990, Eichler:1993,
White:1995}. However, none of the free-free opacity models explain the
radio light curve in a satisfactory manner.  Alternatively, the
variations may be in part due to processes intrinsic to the
WCR. Changes in the stellar separation ($D$) alter the intrinsic
synchrotron luminosity of the WCR ($\propto D^{-1/2}$), and the
free-free absorption and synchrotron self-absorption within the WCR
\citep{Dougherty:2003a}. Furthermore, the impact of the Razin effect
and Coulombic cooling increases as the separation decreases. These
processes are dominant at lower frequencies. Inverse-Compton (IC)
cooling of the shocked gas in the WCR due to the intense ultra-violet
radiation from the stellar components is important at higher
frequencies, and also varies strongly with separation.

We report on high resolution radio observations of WR\thinspace140
obtained with the VLBA that image the WCR at a linear
resolution of a few AU, approximately the stellar separation at
periastron. These observations are used to constrain the orbit of
WR\thinspace140, allowing the first reliable derivation of the orbital
inclination, the longitude of periastron, and a robust distance
estimate. In addition to supporting radiometry from the VLA and
MERLIN, the high resolution observations are essential to constrain
models of the WCR in WR\thinspace140.

\section{Observations}
\begin{figure*}
%\hrule
\vspace{21.4cm}
\includegraphics{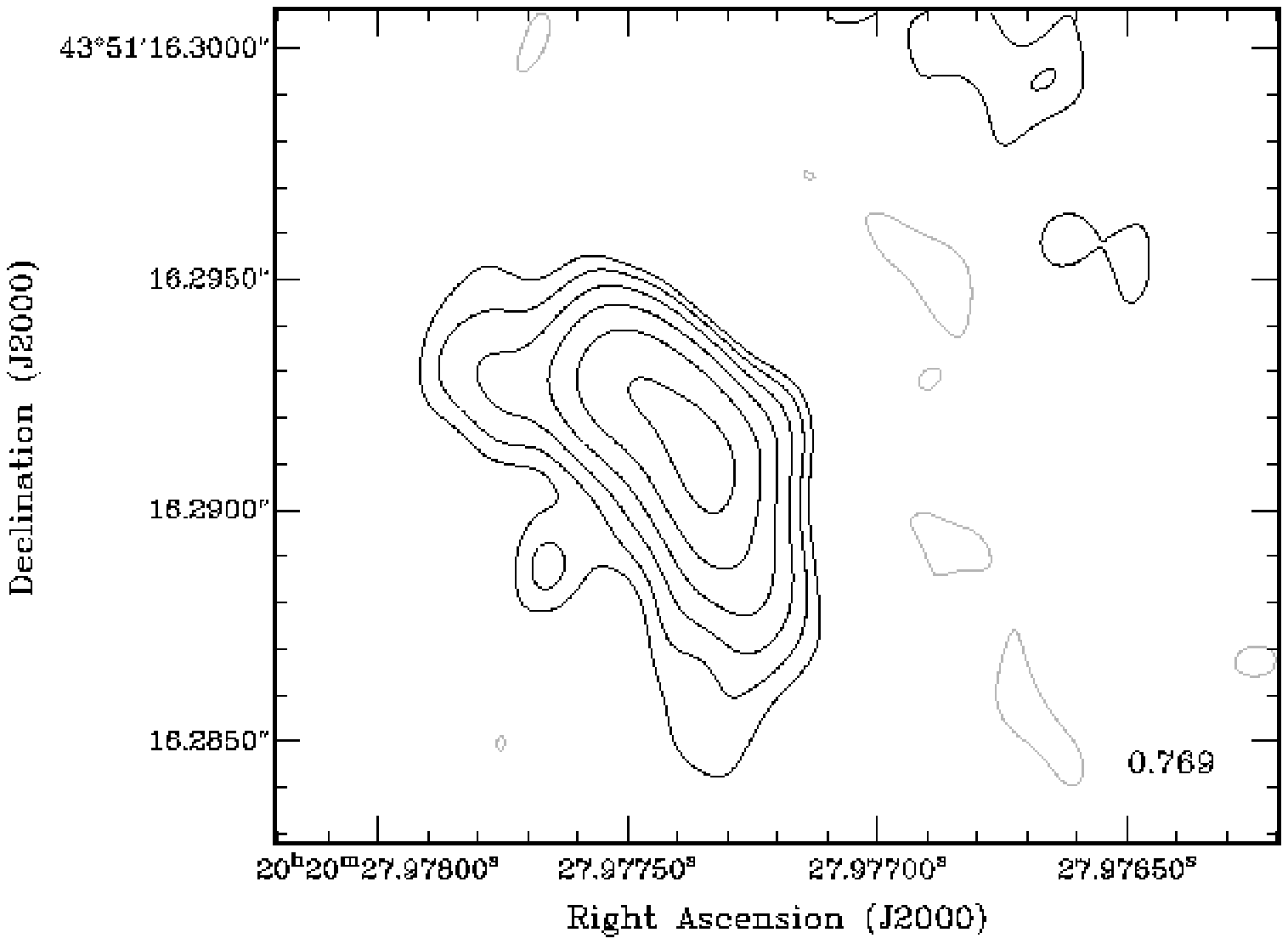}
\includegraphics{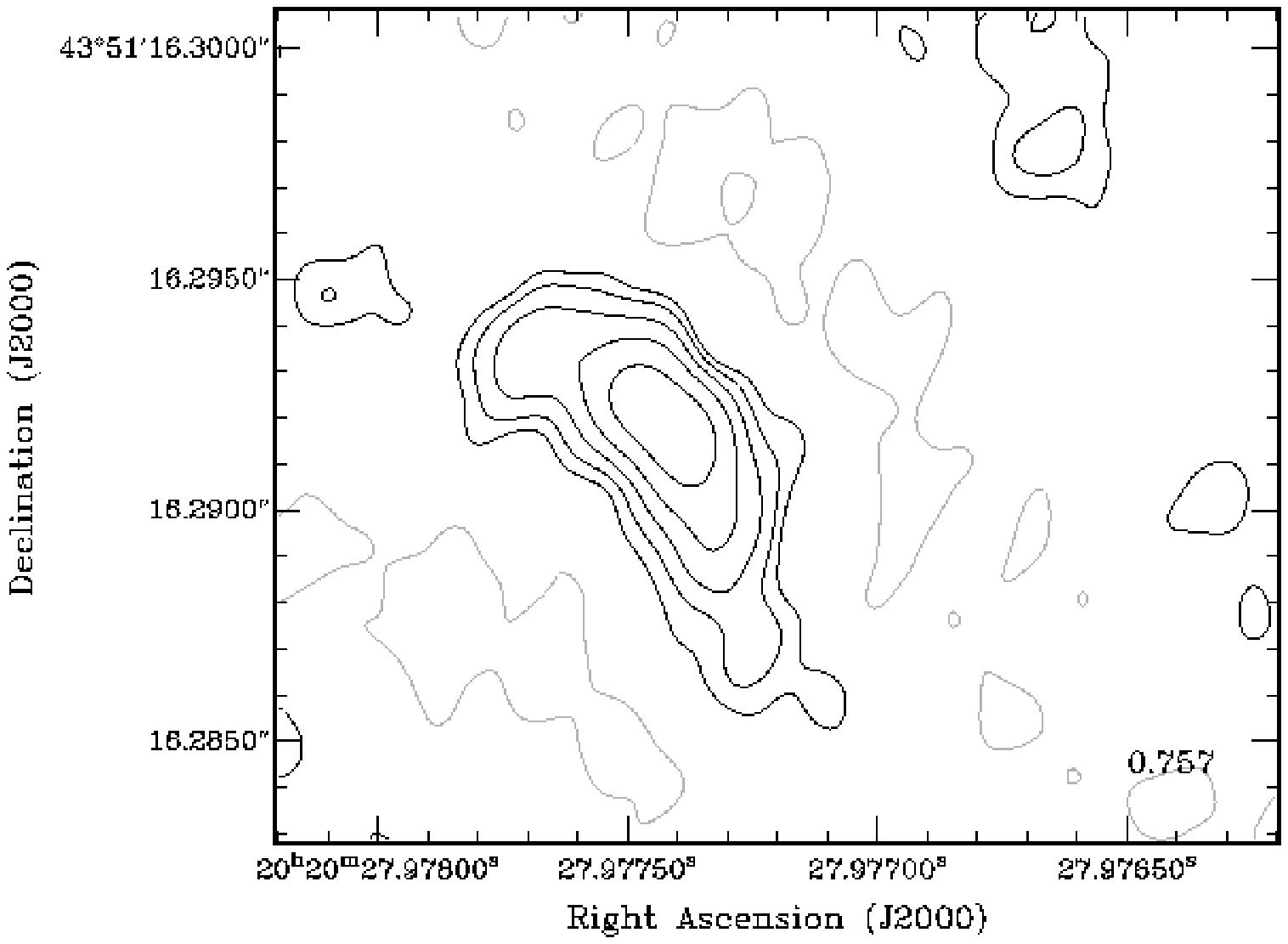}
\includegraphics{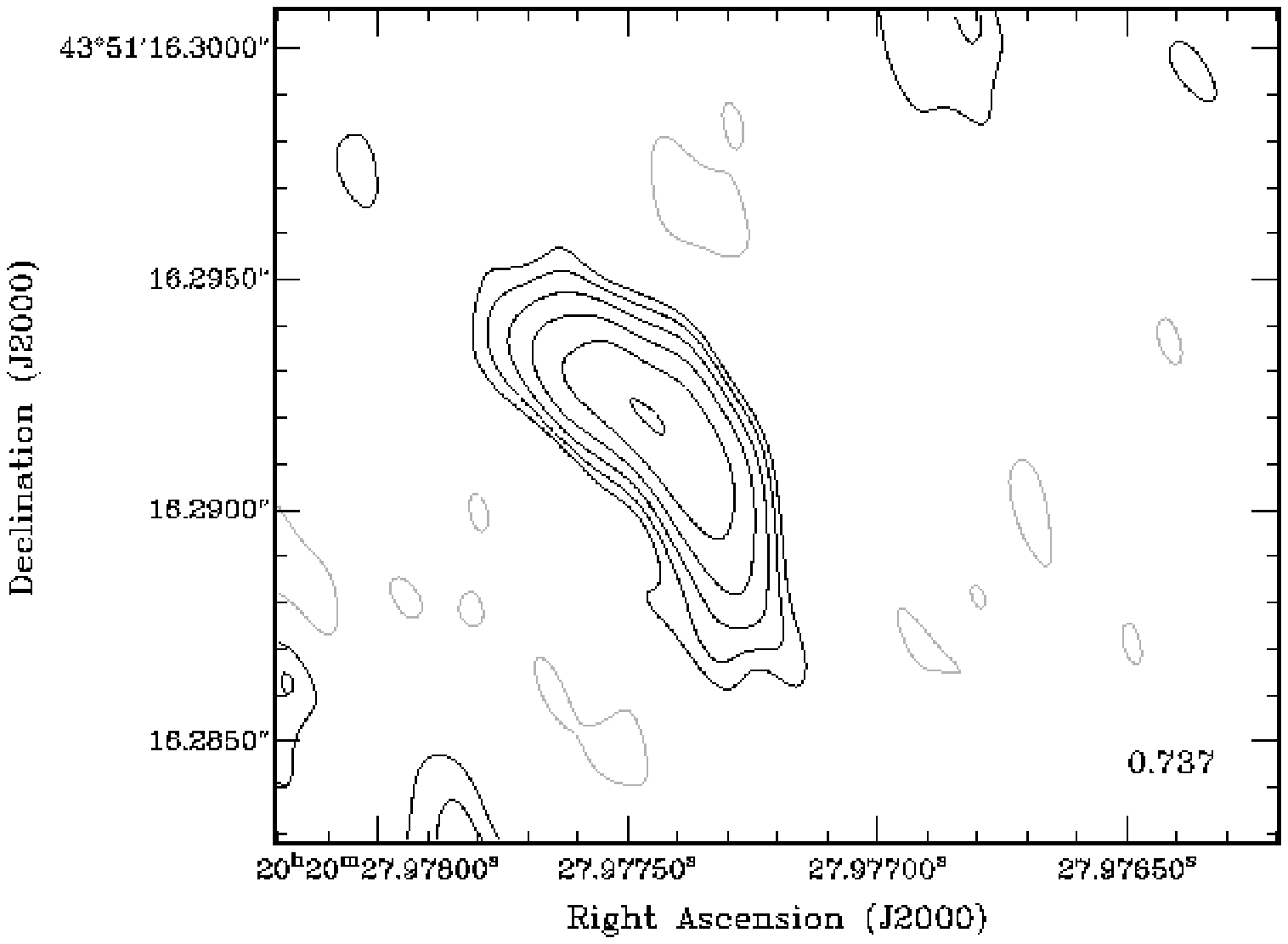}
\includegraphics{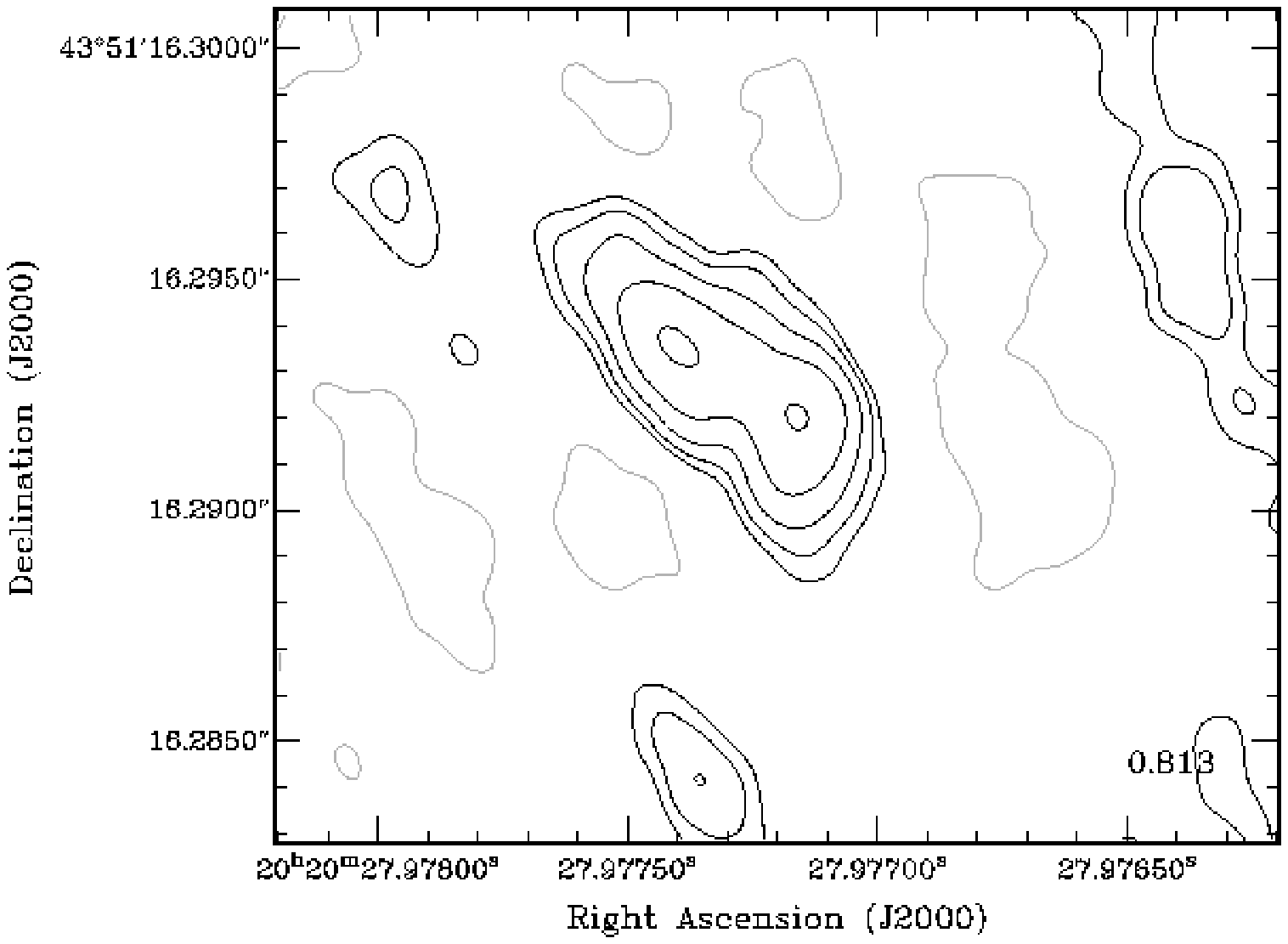}
\includegraphics{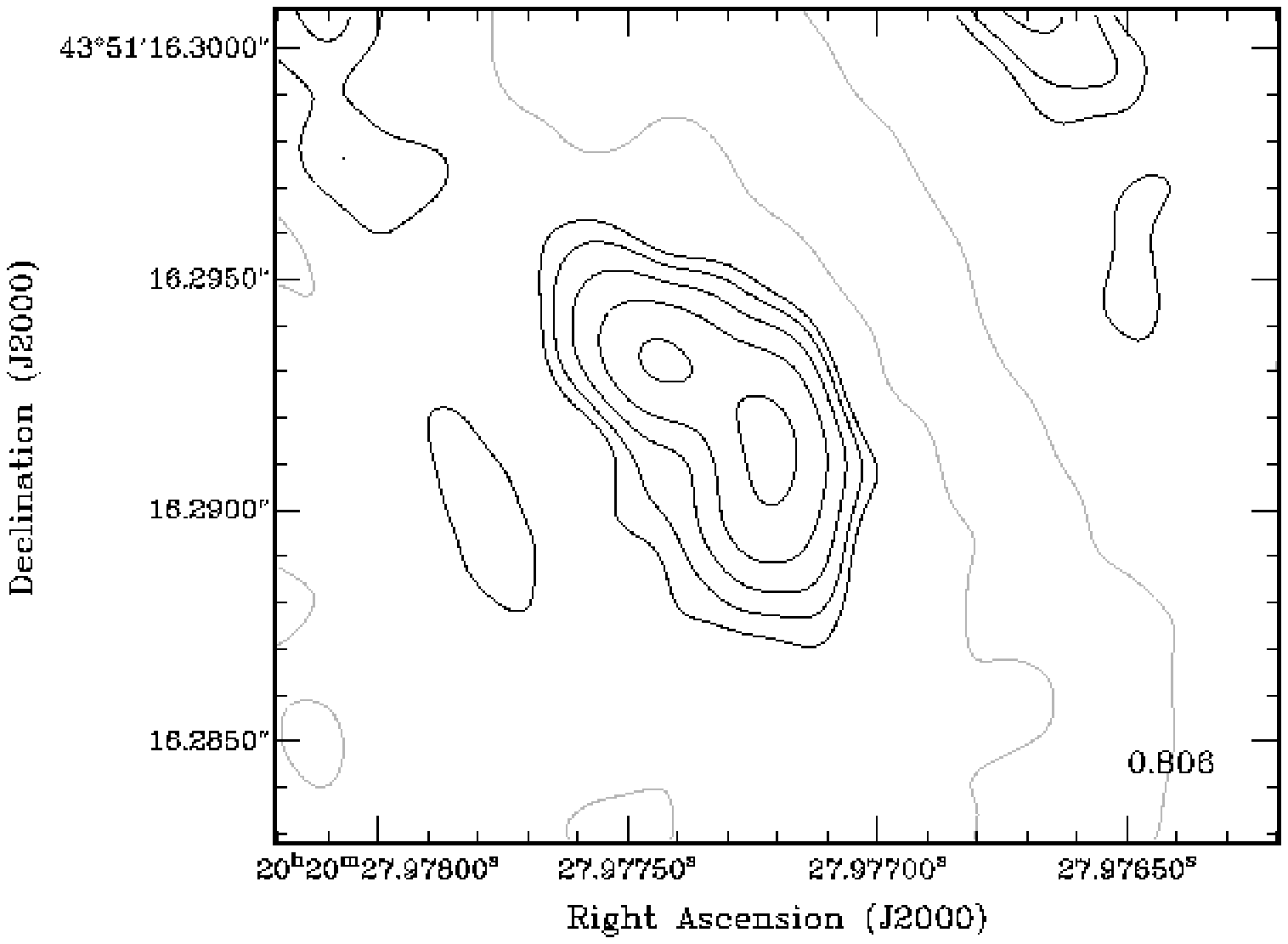}
\includegraphics{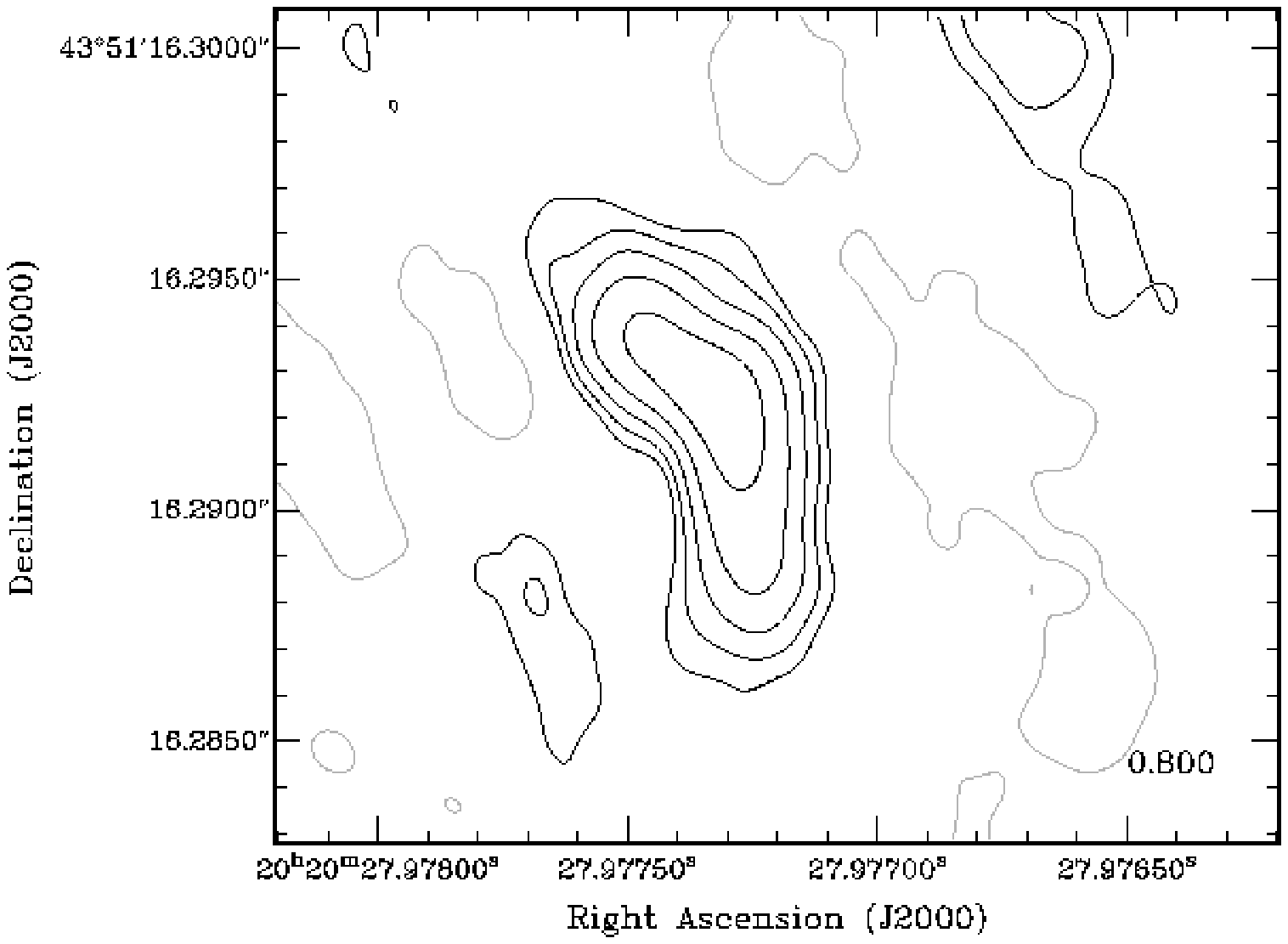}
\includegraphics{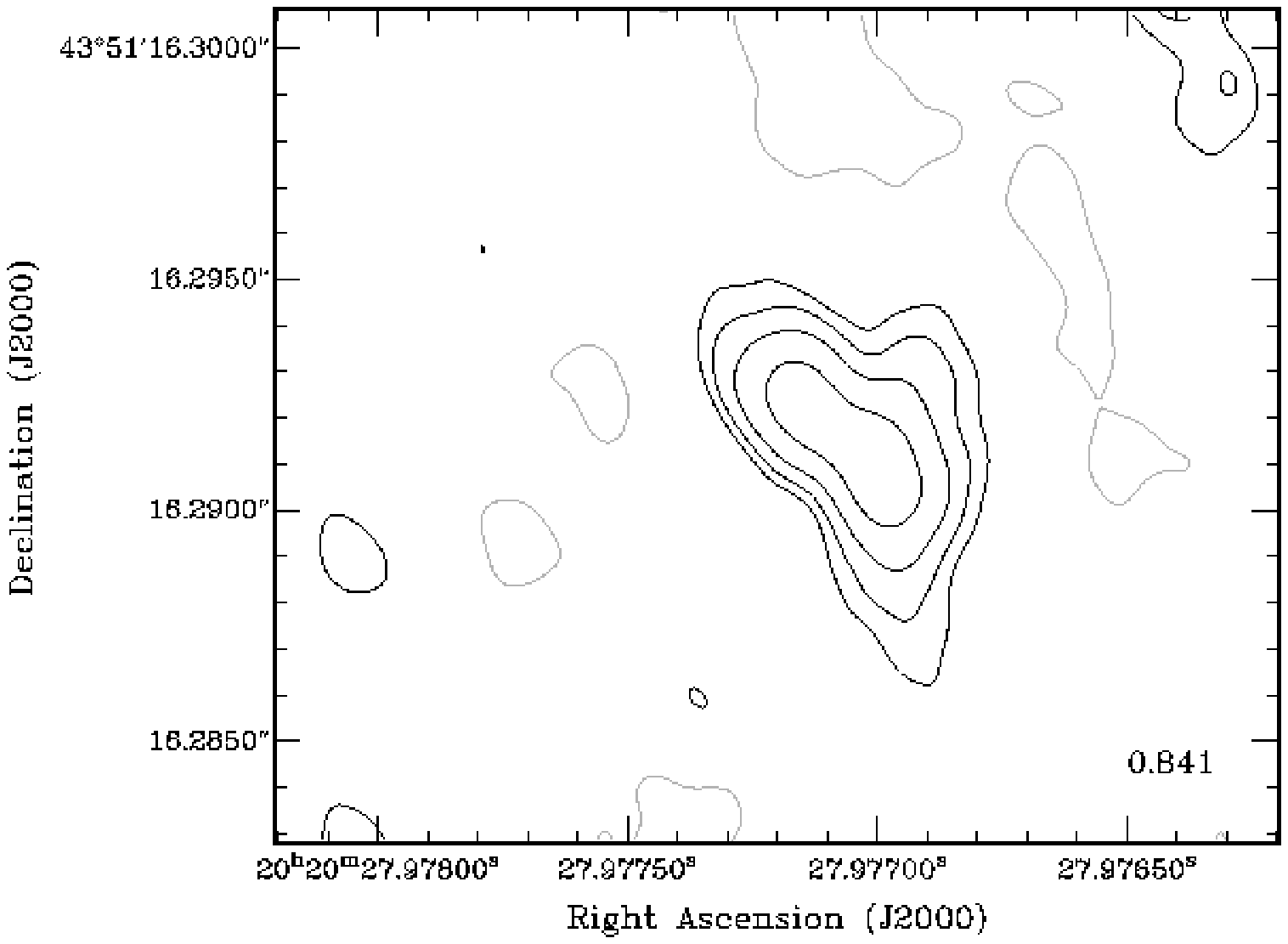}
\includegraphics{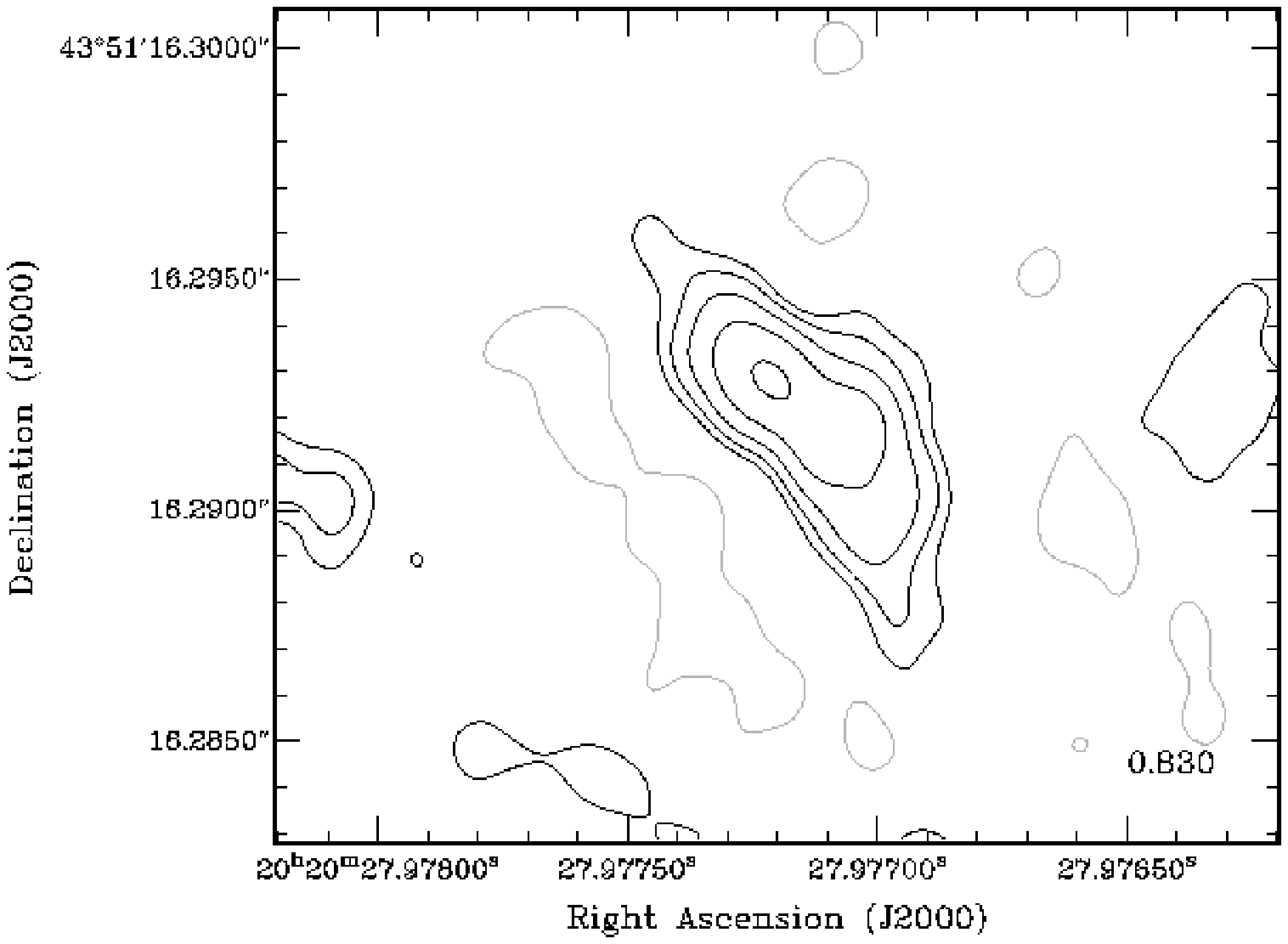}
\includegraphics{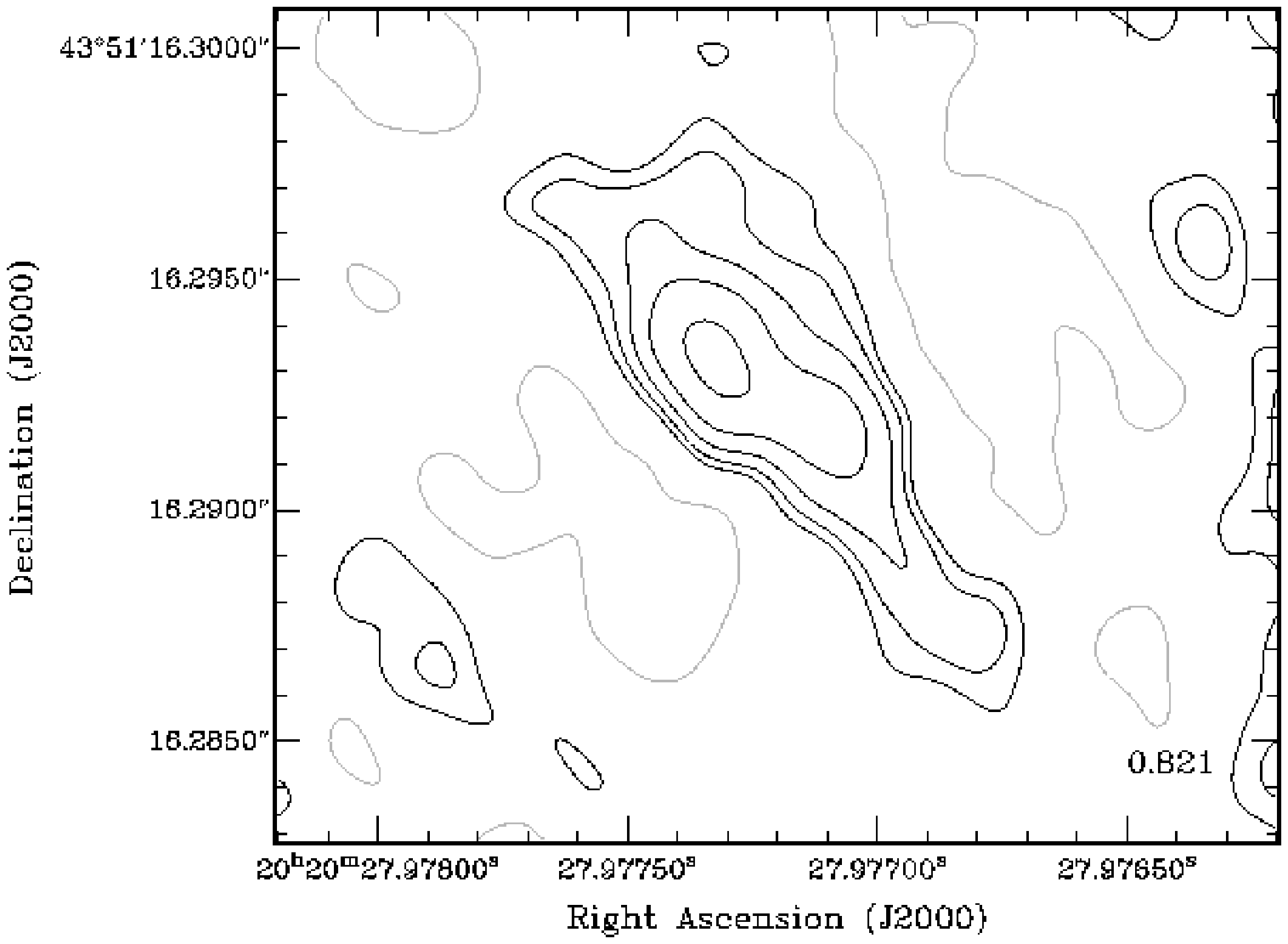}
\includegraphics{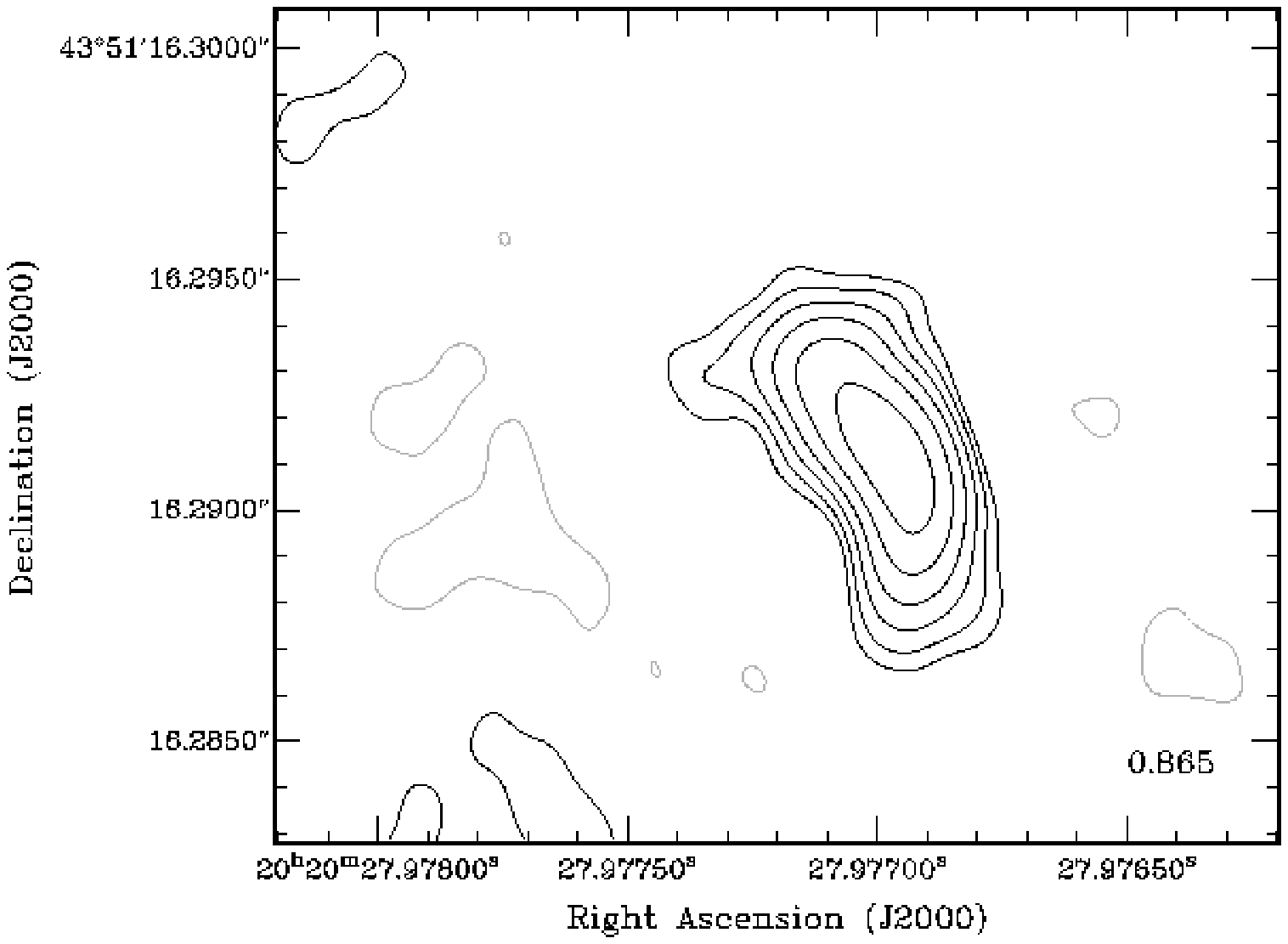}
\includegraphics{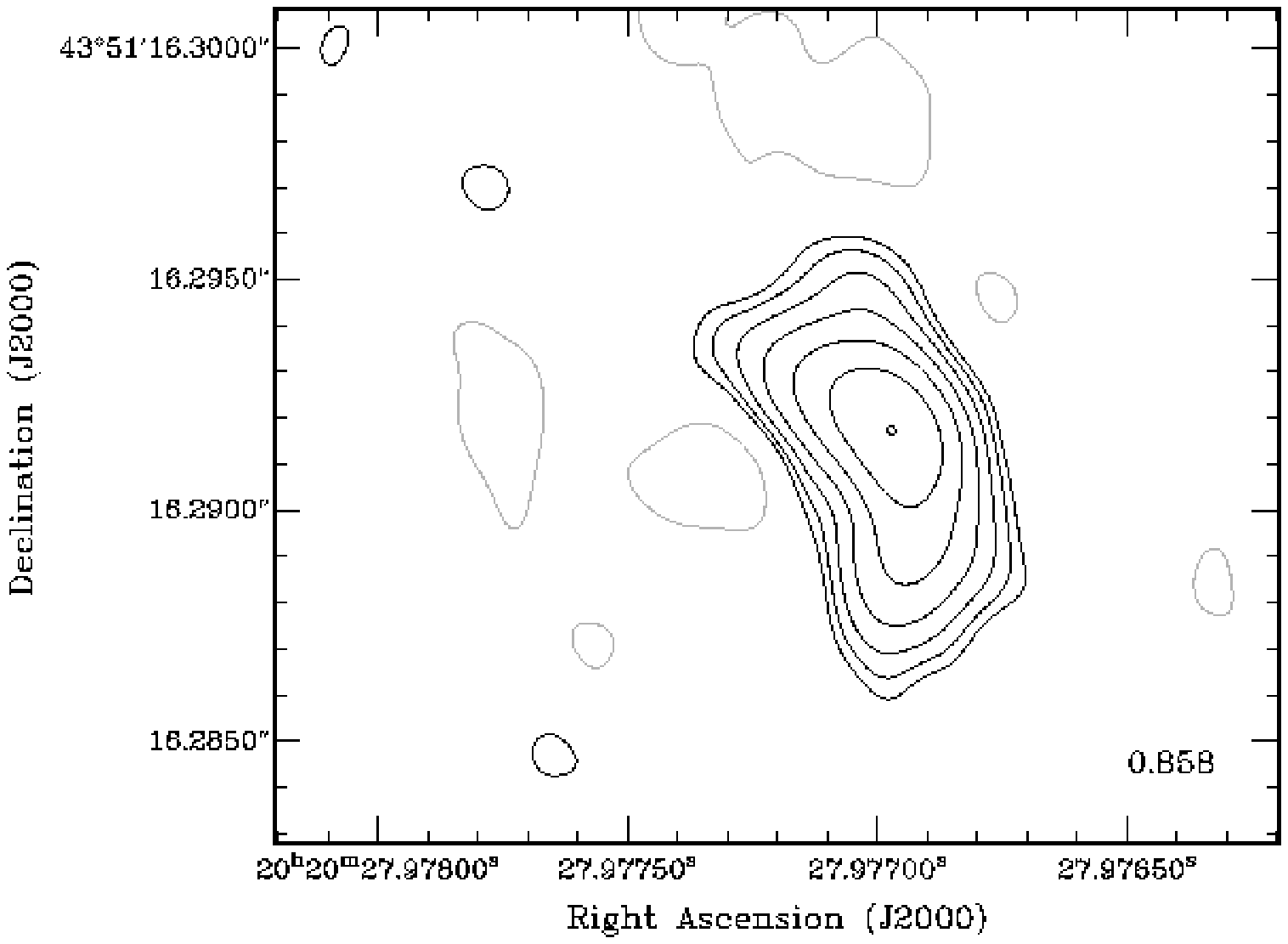}
\includegraphics{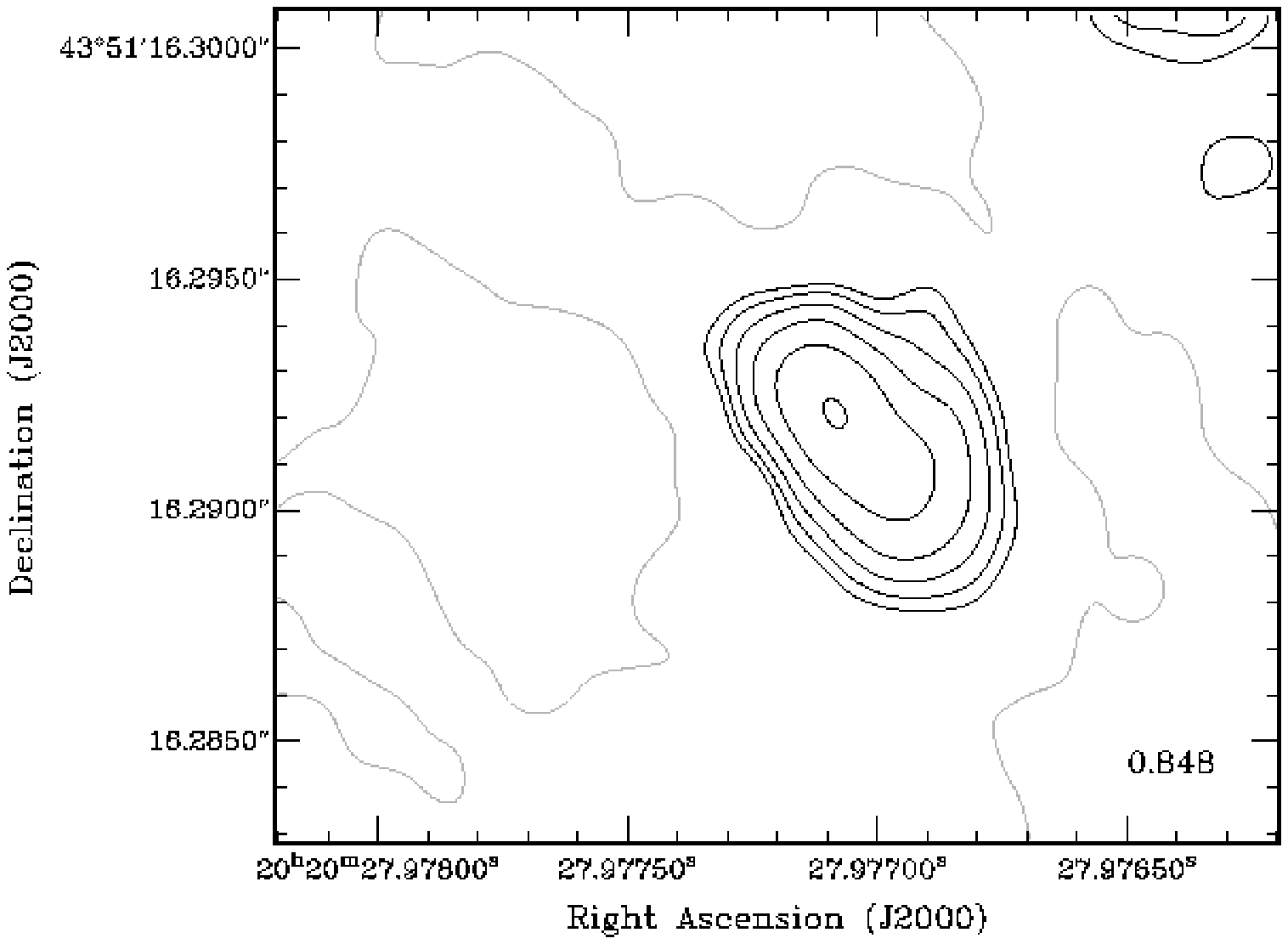}
\includegraphics{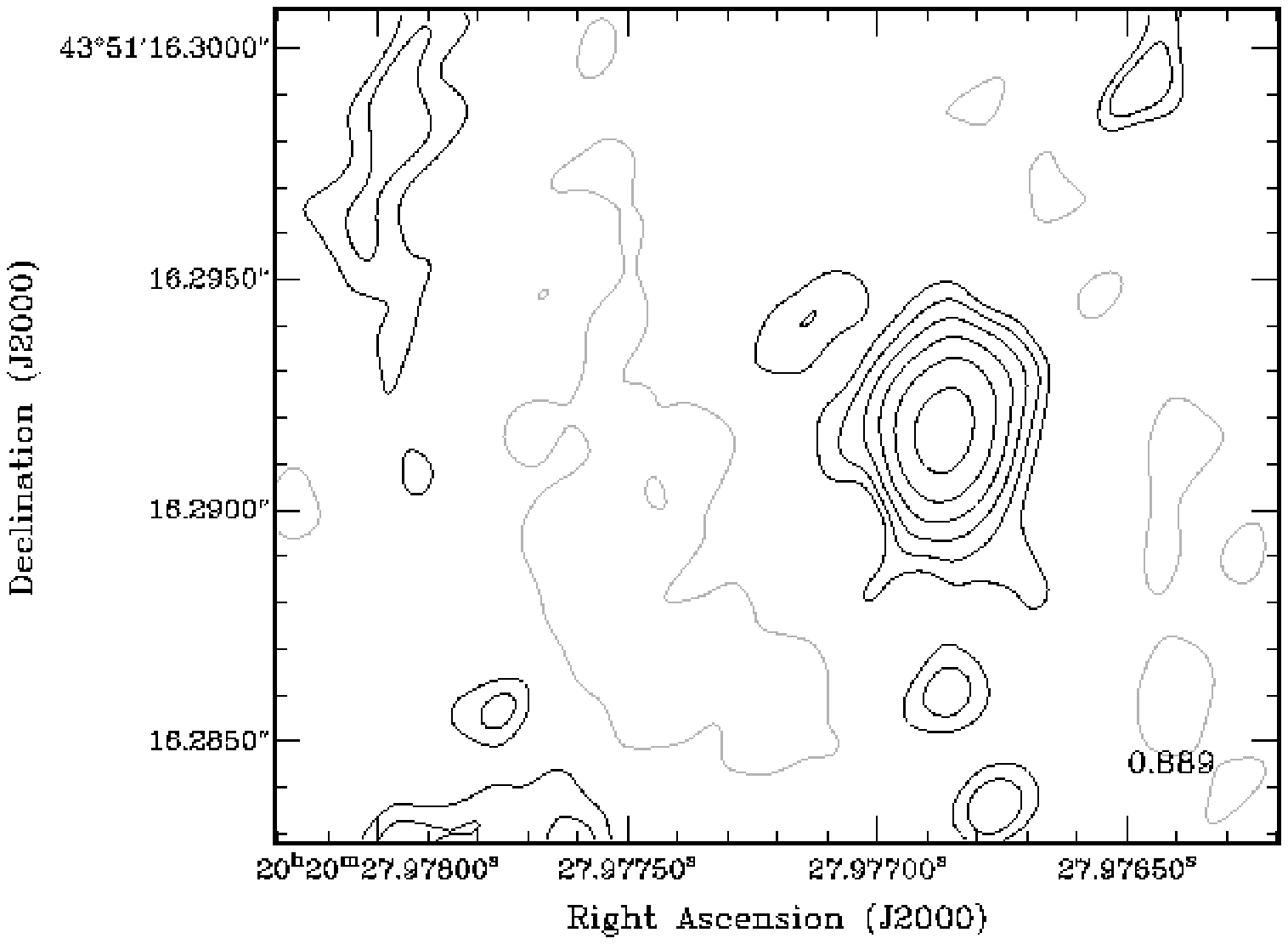}
\includegraphics{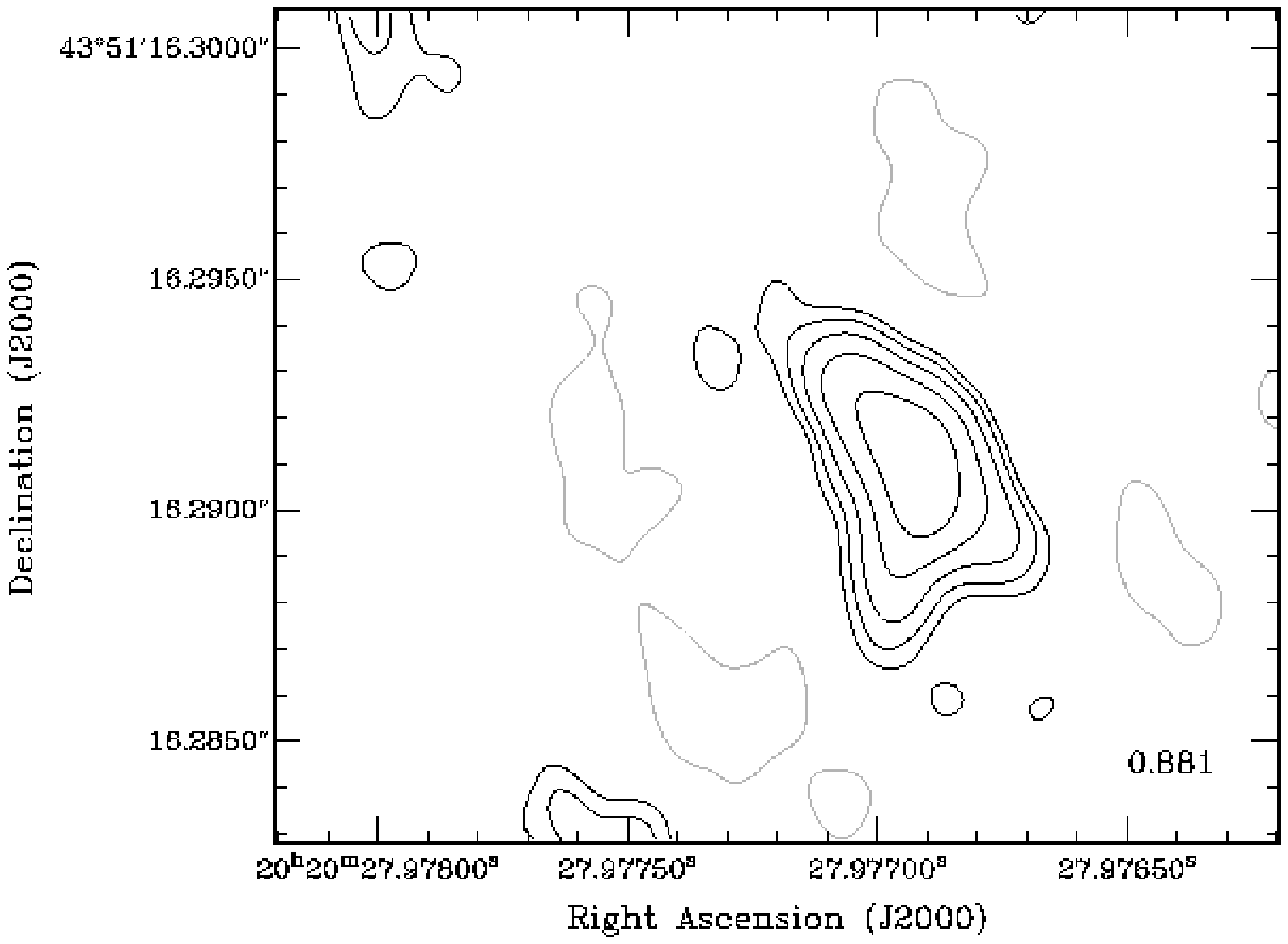}
\includegraphics{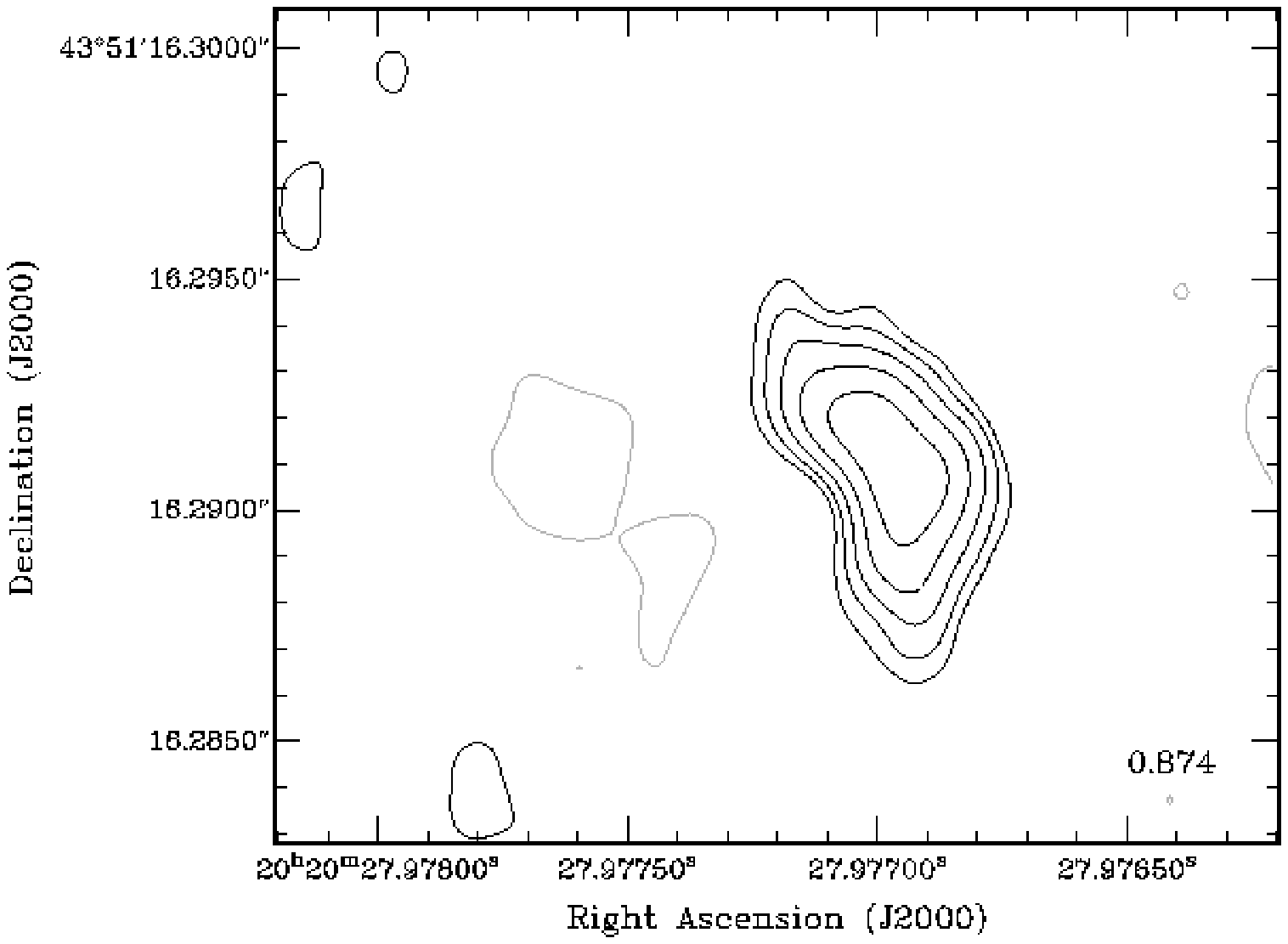}
%\hrule
\caption{8.4-GHz VLBA observations of WR\thinspace140 from orbital
phase 0.737 through 0.889. The contour levels are
-1, 1, 1.6, 2.6, 4.1, 6.5, 10.5$\rho$ where $\rho=220~\mu$Jy~beam$^{-1}$. Note
that $\rho$ is a common contour scale factor between all the epochs to
enable easy comparison of each image, and is {\em not} the rms
background level which is between 0.06 and 0.22~mJy~beam$^{-1}$. The
greyscale extends from -0.4 (white) to 2~mJy~beam$^{-1}$ (black).  The
contour levels and greyscale are identical in each image. The typical
beam size in the images is
$2.0\times1.5$~mas$^2$.\label{fig:xband_vlba}}
\end{figure*}
\addtocounter{figure}{-1}
\begin{figure*}[t]
%\hrule
\vspace{13cm}
\includegraphics{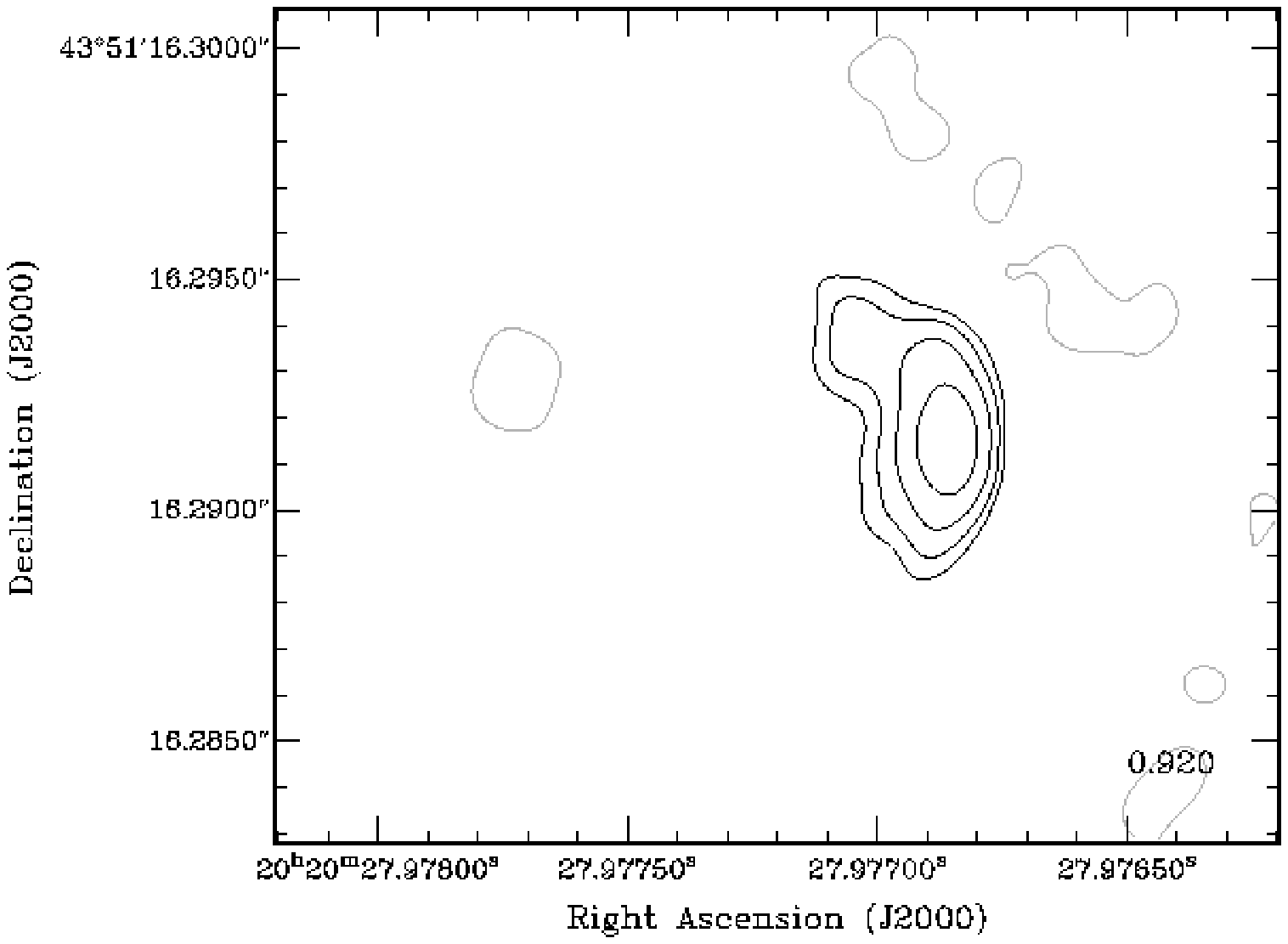}
\includegraphics{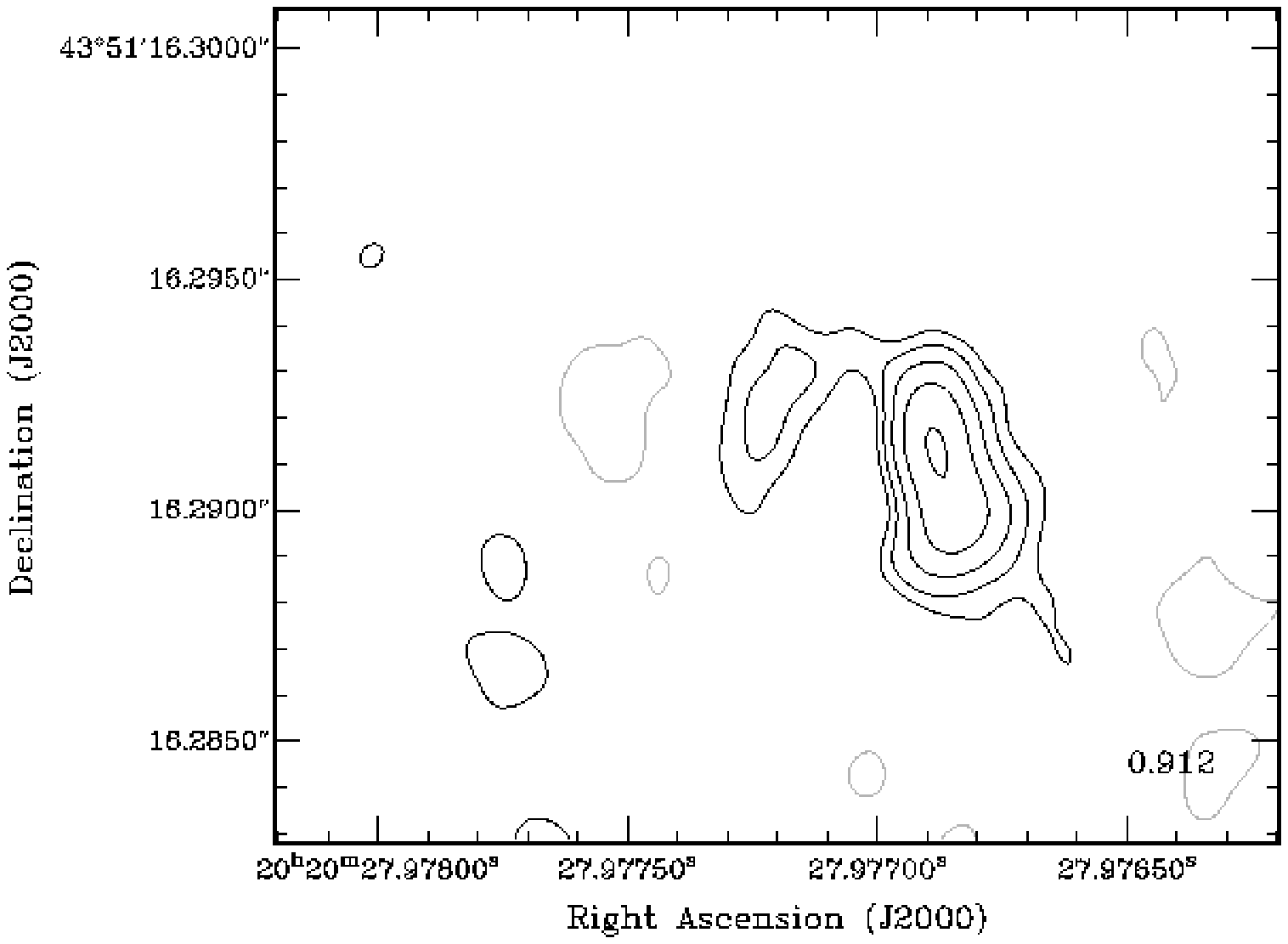}
\includegraphics{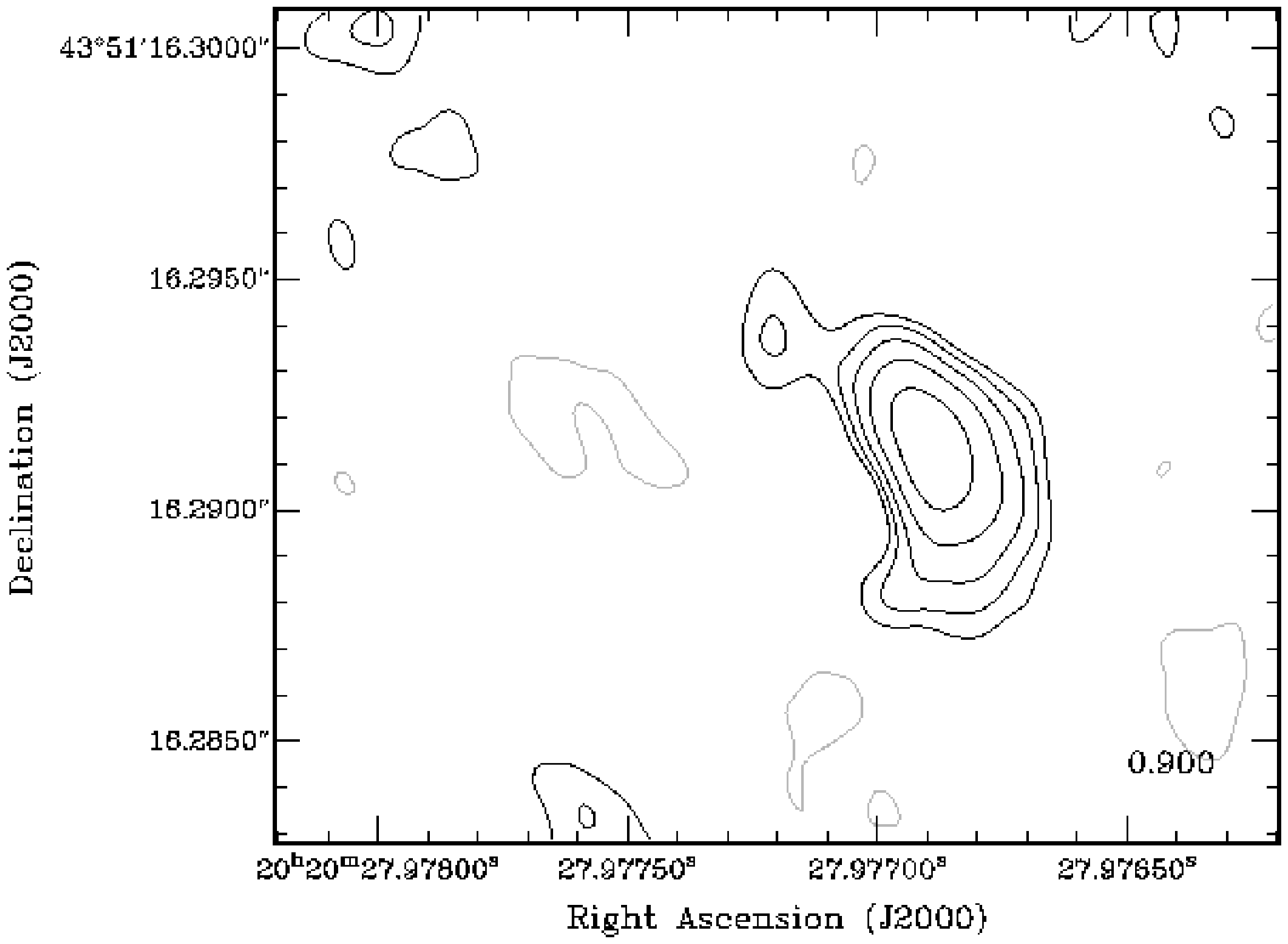}
\includegraphics{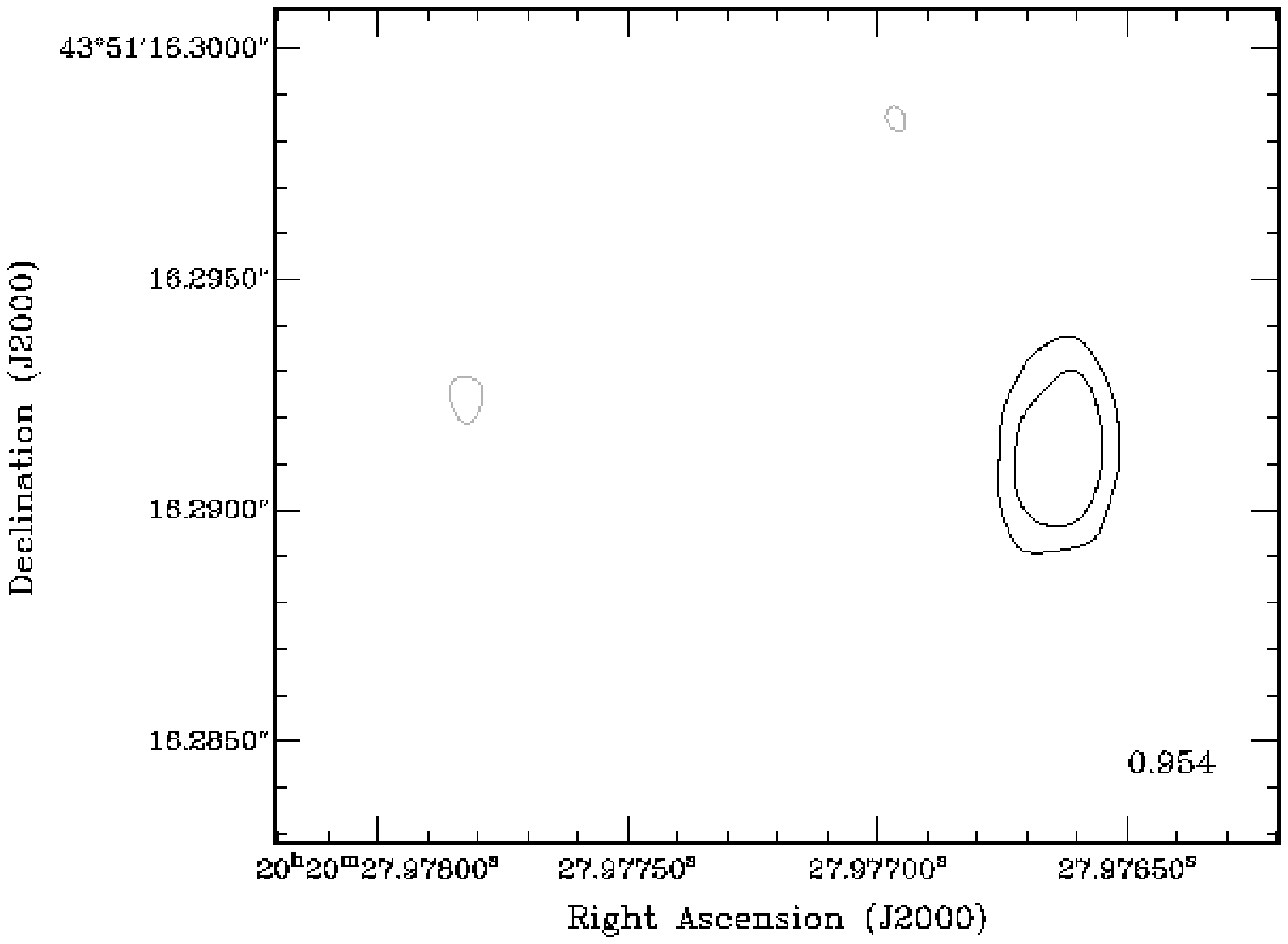}
\includegraphics{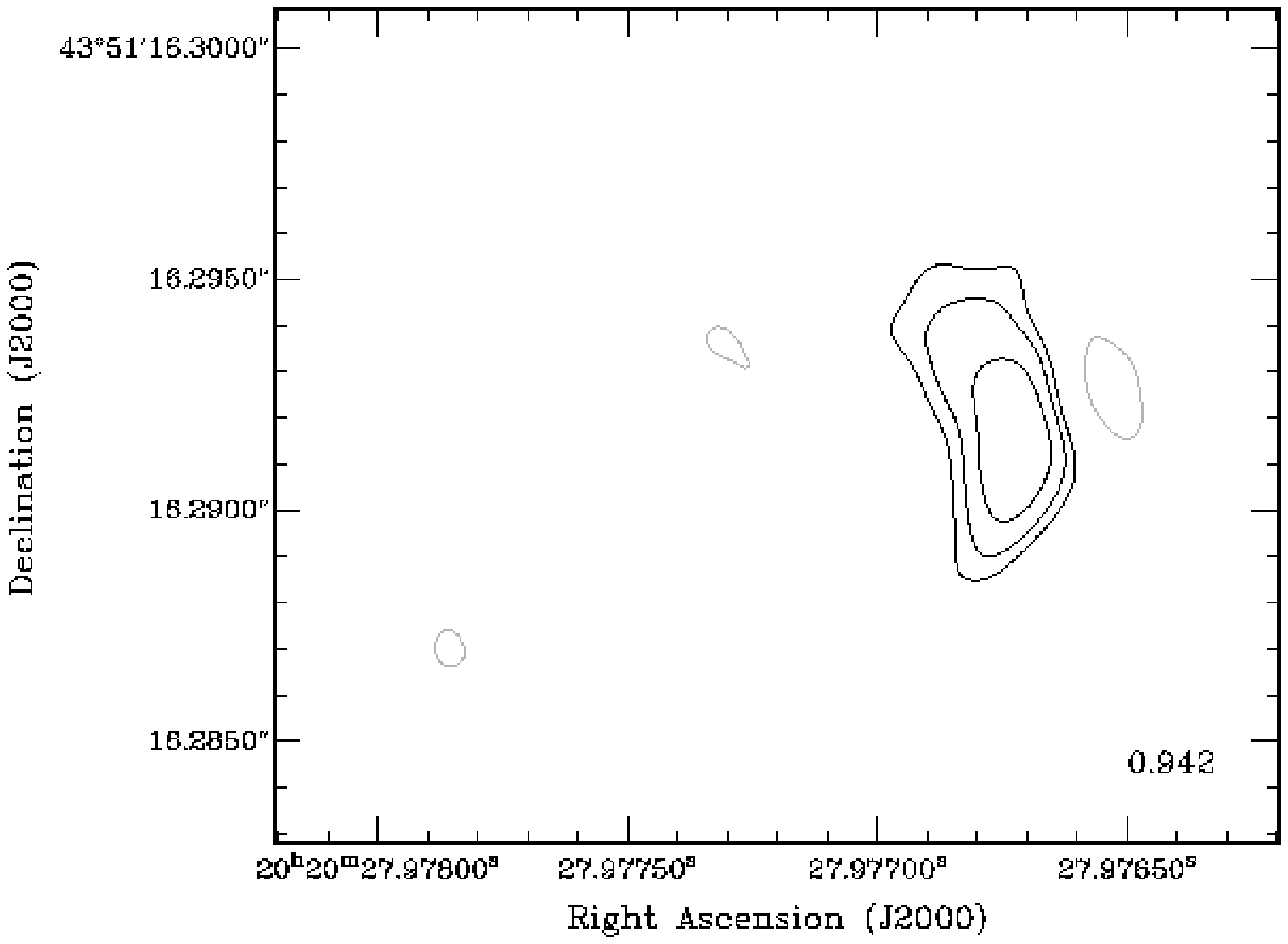}
\includegraphics{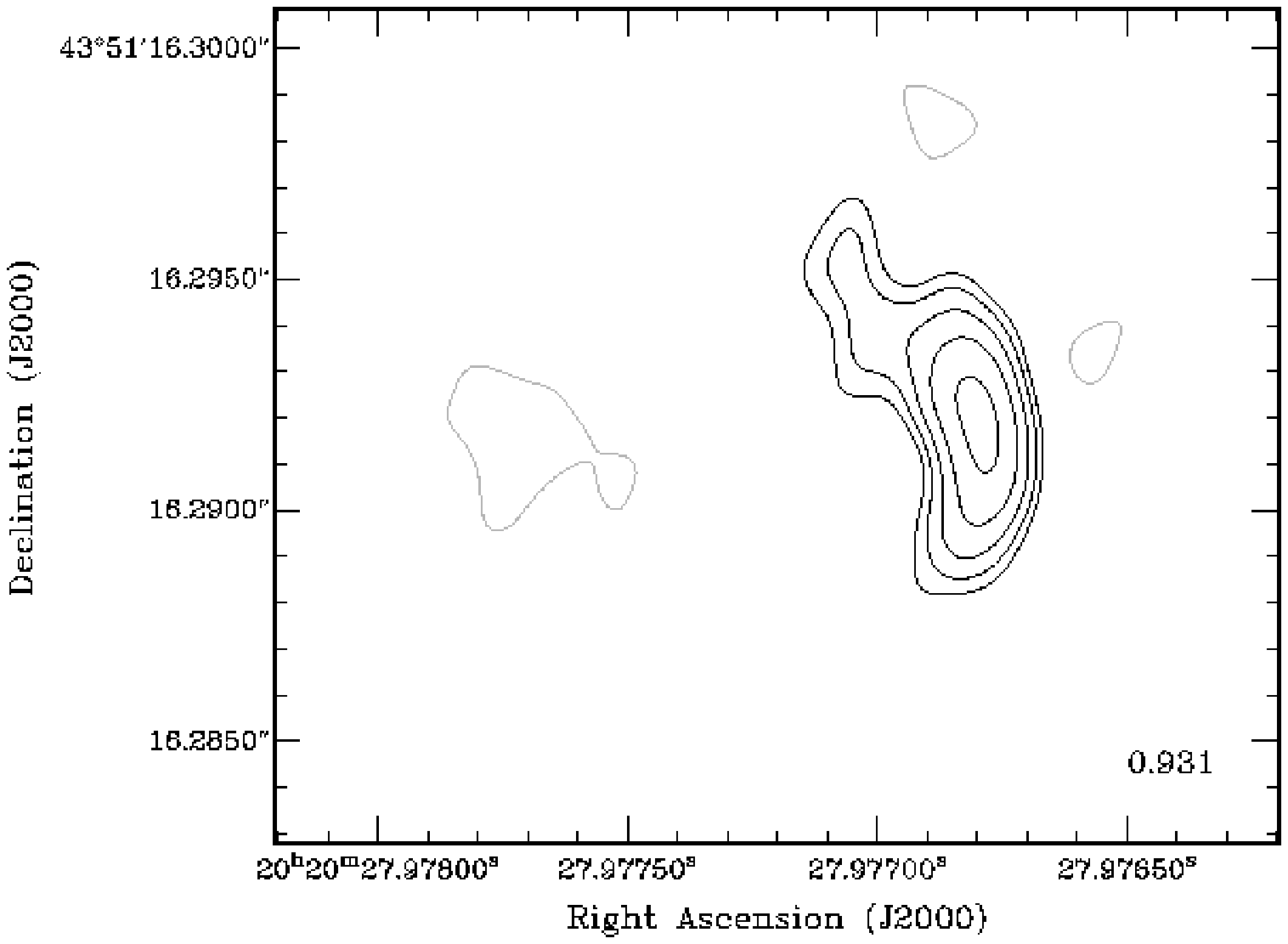}
\includegraphics{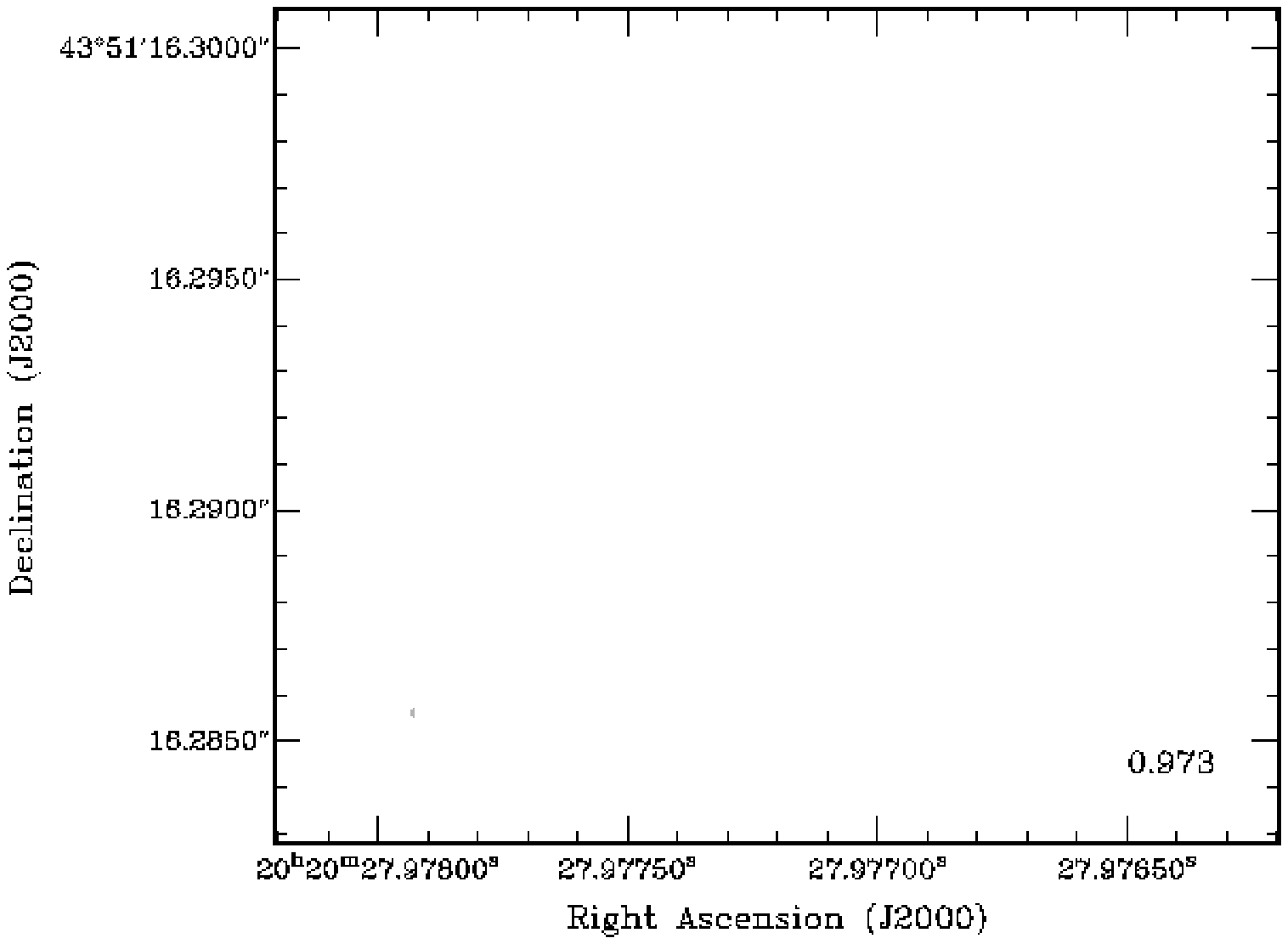}
\includegraphics{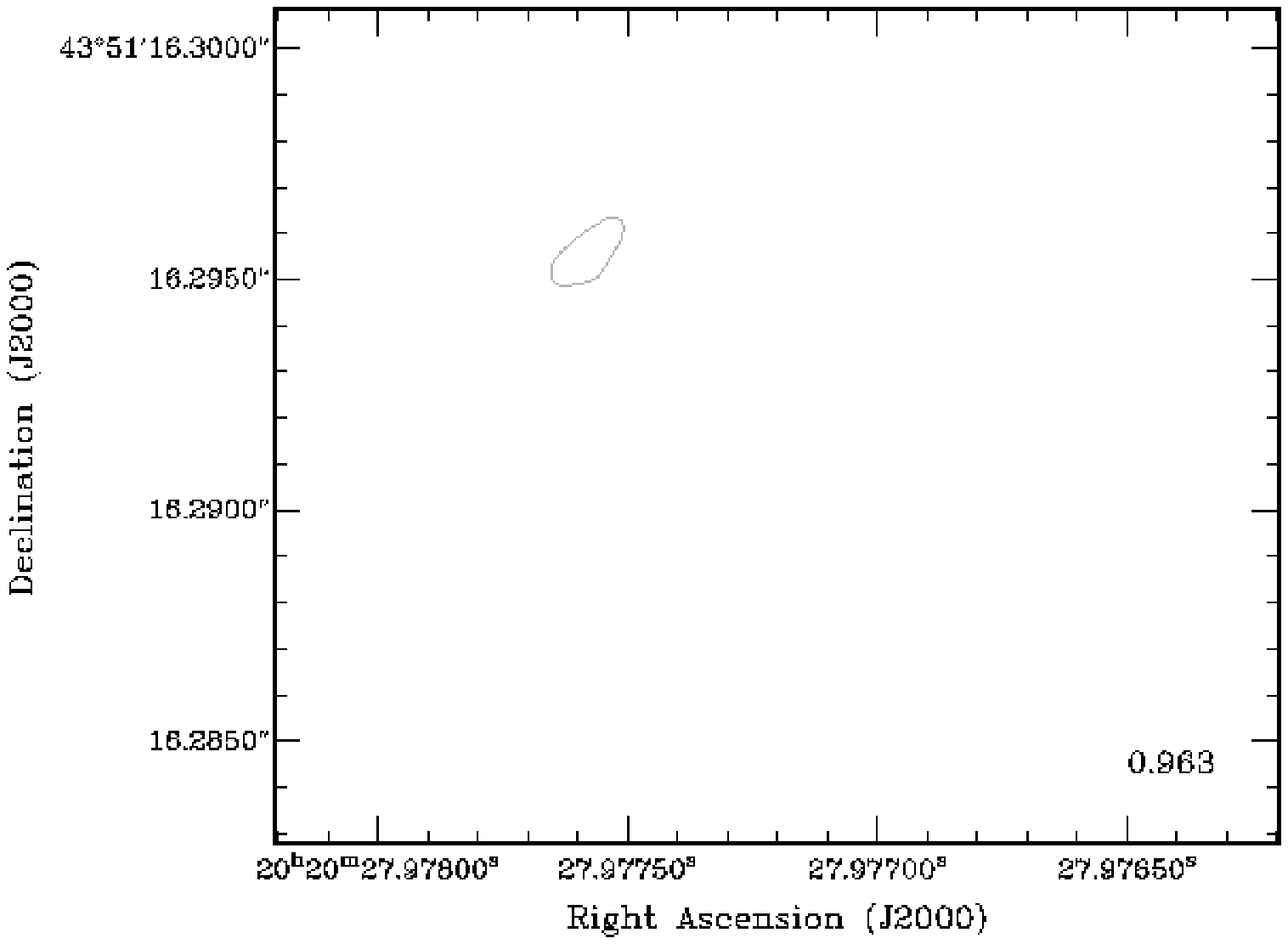}
%\hrule
\caption{(continued) 8.4-GHz VLBA observations of WR\thinspace140 from
orbital phase 0.900 through 0.973. Contour levels and grey scales are
the same as Fig.~\ref{fig:xband_vlba}a.  }
\end{figure*}

\begin{figure*}
%\hrule
\vspace{4.55cm}
\includegraphics{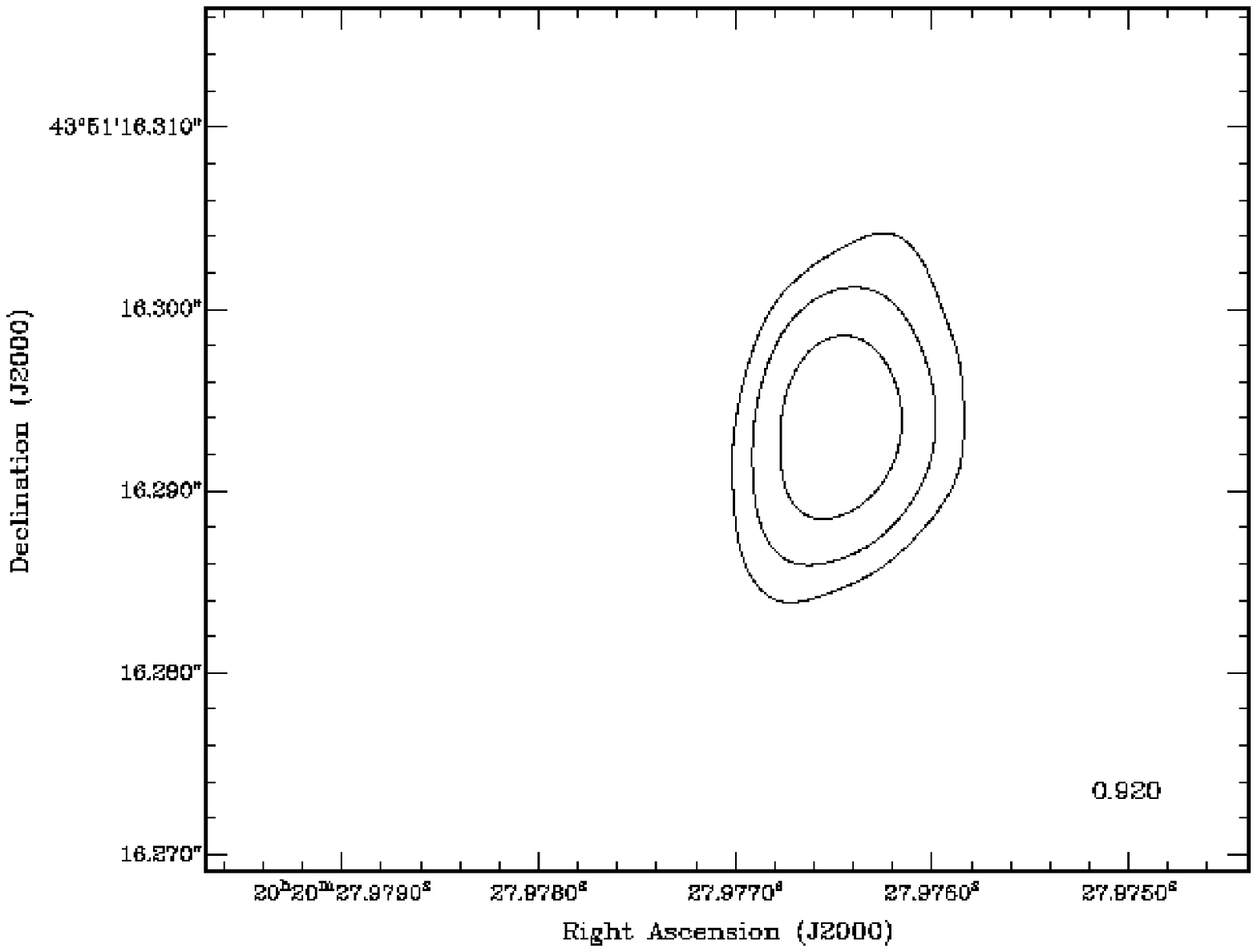}
\includegraphics{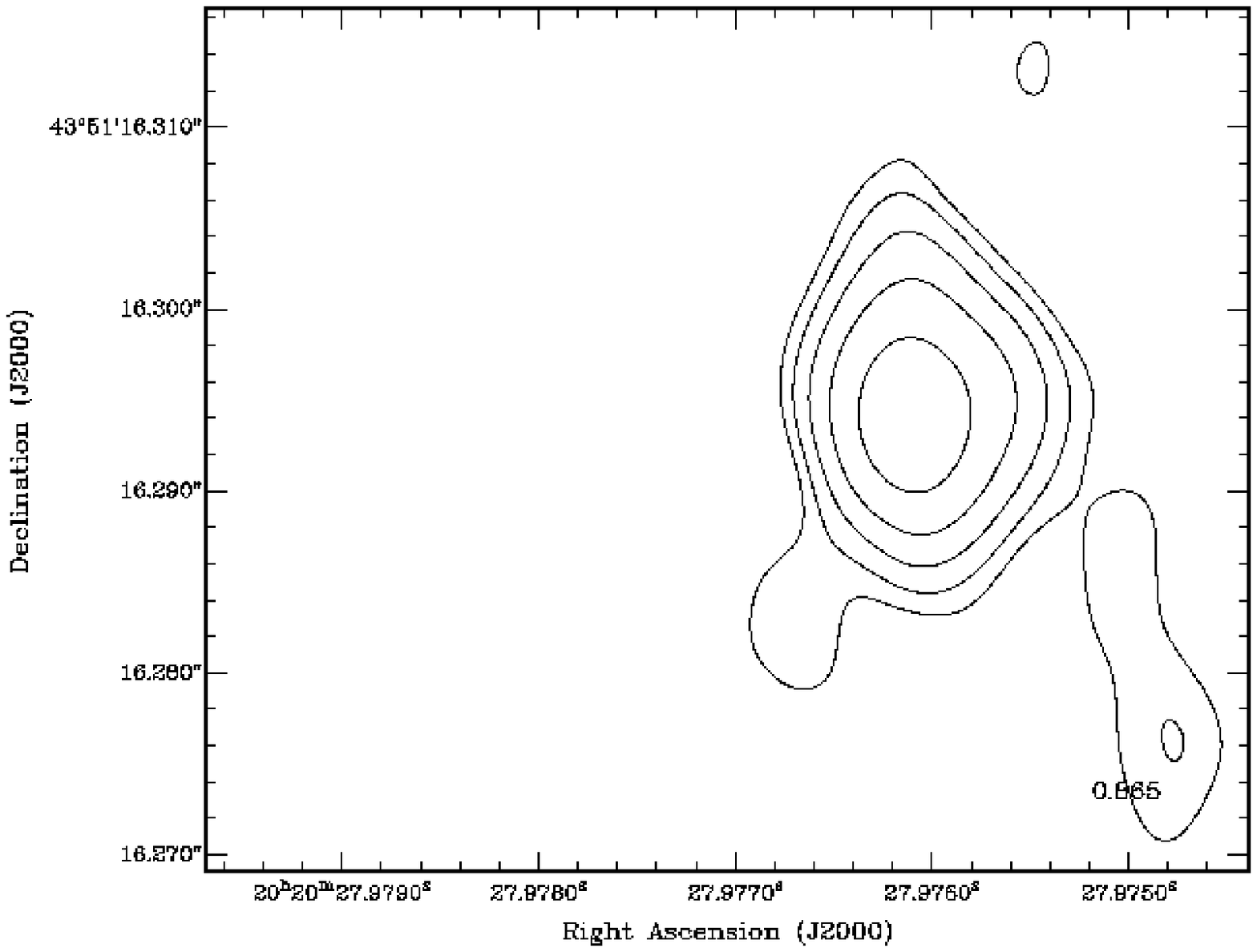}
\includegraphics{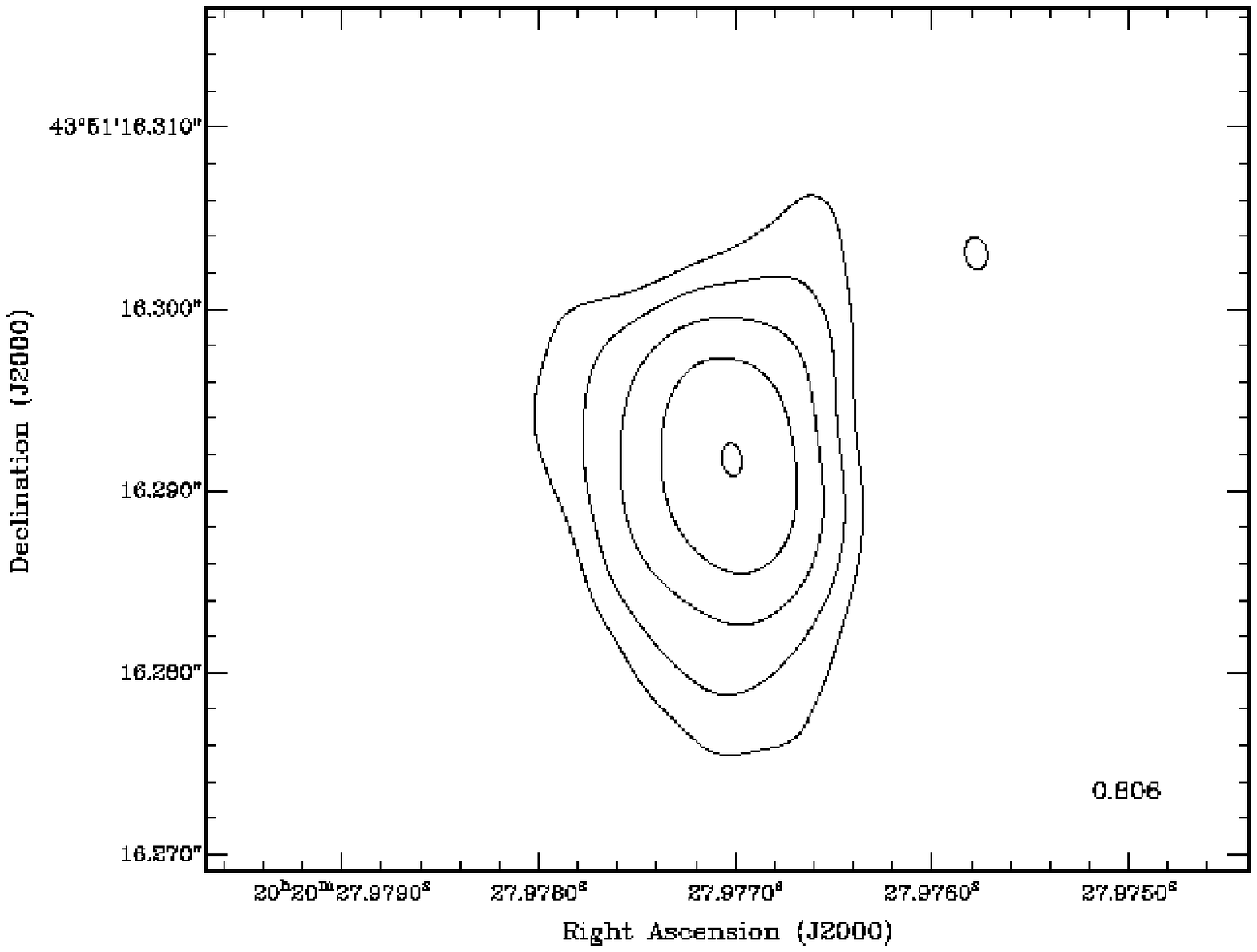}
%\hrule
\caption{Examples of 1.7-GHz VLBA observations of WR\thinspace140 at
orbit phases 0.806, 0.865, and 0.920. The contour levels are -1, 1,
1.6, 2.6, 4.1, 6.5$\rho$ where $\rho=0.9$~mJy.  The greyscale
extends from -0.1 (white) to 8.3~mJy~beam$^{-1}$ (black).  The contour
levels and greyscale are identical in each image. The typical beam
size in the images is $10\times5$~mas$^2$. \label{fig:lband_vlba} }
\end{figure*}

\subsection{VLBA observations}
\label{sec:vlba_obs}

A 23-epoch campaign of observations of WR\thinspace140 at both 8421
MHz (X band) and 1667 MHz (L band) was carried out using the VLBA
\citep{Napier:1995}. The campaign was started on 1999 January 04 near
the peak of radio emission around orbital phase 0.75. Observations
continued until 2000 November 18 ($\phi=0.97$) when the radio emission
had declined to close to radio minimum (see Fig:\ref{fig:light_curve}
and could no longer be detected by the VLBA.

Data were received in dual circular polarizations, and acquired from
four contiguous 8-MHz bands using four-level (2 bit) sampling for each
polarization, providing a 32 MHz total bandwidth. The observations
were performed using phase referencing \citep[see][]{Beasley:1995}
where regular observations of a nearby extra-galactic background
source were made to derive residual antenna-based instrumental and
atmospheric phase errors, which were then interpolated to correct the
antenna gains during observations of WR\thinspace140. This technique
allows the detection of weak emission by enabling direct Fourier
inversion of the visibility data for imaging of WR\thinspace140, along
with precise astrometric referencing of the images. The phase
calibrator was J2002+456, separated from WR\thinspace140 by
$2.44\deg$. A switching cycle of 90 seconds on a calibrator and 120
seconds on the target source was used, giving $\sim3$~hrs on-source
time for WR\thinspace140 at each epoch at each frequency.  The
absolute flux scale of the observations was based on internal noise
source calibration.  The data were correlated with the VLBA correlator
in Socorro, New Mexico, and data calibration and imaging was done
using standard techniques in the NRAO {\sc aips} software package.

The resulting phase-referenced images at 8.4-GHz for each observation
epochs are shown in Fig.~\ref{fig:xband_vlba}, where the rms
uncertainty of the image background is between 0.06 and 0.22
mJy~beam$^{-1}$. The nominal resolution (FWHM) of the synthesized beam
is $2.0\times1.5$~mas$^2$, which at a distance of 1.85 kpc (see
\textsection{3.2}) gives a linear resolution of 3-4 AU, much smaller
than the separation of the stellar components in WR\thinspace140 for
the bulk of the orbit except close to periastron.

\begin{deluxetable*}{ccccccccccc}
\tabletypesize{\scriptsize}
\tablecaption{Positions and fluxes from the VLBA observations\label{tab:vlba_obs}}
\setlength{\tabcolsep}{0.02in} 
\tablewidth{0pt}
\tablehead{
\colhead{Obs. Code} & 
\colhead{Obs. Date} & 
\colhead{Phase\tablenotemark{a}} & 
\multicolumn{5}{c}{8.4 GHz data} &
\multicolumn{3}{c}{1.7 GHz data} \\ %\cline{4-8}\cline{9-11}
&&&
\colhead{$\Delta$$\alpha$\tablenotemark{b,c}} & 
\colhead{$\Delta$$\delta$\tablenotemark{b,c}} & 
\colhead{Flux \tablenotemark{c}} & 
\colhead{PA\tablenotemark{d}} &
\colhead{$\Theta$\tablenotemark{e}} &
\colhead{$\Delta$$\alpha$\tablenotemark{b,c}} & 
\colhead{$\Delta$$\delta$\tablenotemark{b,c}} & 
\colhead{Flux \tablenotemark{c}} \\
&&&
\colhead{(ms)} &\colhead{(mas)} &\colhead{(mJy)}&\colhead{($\deg$)} &
\colhead{($\deg$)} &\colhead{(ms)} &\colhead{(mas)} &\colhead{(mJy)} 
}
\startdata
BB102A & 1999/01/04 &0.737 &  0     & 0    &10.5 & $47\pm4$  &73 &-0.889 &4.45  &12.4\\
BB102B & 1999/03/03 &0.757 & -0.003 & 0.01 &10.2 & $47\pm4$  &77 &-0.423 &-1.33 &8.3 \\
BB110A & 1999/04/08 &0.769 & -0.045 &-0.34 &10.3 & $56\pm5$  &70 &-0.428 &0.86  &14.0\\
BB109A & 1999/07/05 &0.800 & -0.106 & 0.46 &10.4 & $53\pm4$  &71 &-0.526 &0.37  &10.4\\
BB109B & 1999/07/23 &0.806 & -0.126 & 0.19 &8.3  & $48\pm15$ &63 &-0.394 &-0.56 &14.6\\
BB109C & 1999/08/12 &0.813 & -0.188 & 0.77 &10.2 & $53\pm11$ &66 &-0.528 &1.84  &16.8\\
BB109D & 1999/09/06 &0.821 & -0.184 & 1.44 &11.0 & $47\pm15$ &85 &-0.601 &2.70  &18.0\\
BB109E & 1999/10/02 &0.830 & -0.232 & 0.88 &8.8  & $50\pm15$ &77 &-0.699 &0.32  &16.3\\
BB109F & 1999/11/01 &0.841 & -0.400 &-0.26 &8.2  & $54\pm10$ &73 &-0.773 &0.23  &9.5\\
BB117A & 1999/11/21 &0.848 & -0.416 &-0.14 &8.8  & $55\pm11$ &88 &-     & -    &\tablenotemark{f}\\
BB117B & 1999/12/22 &0.858 & -0.464 &-0.30 &10.5 & $55\pm4$  &68 &-1.215 &4.75  &16.5\\
BB117C & 2000/01/11 &0.865 & -0.450 &-0.42 &9.3  & $59\pm4$  &74 &-1.374 &2.96  &17.3\\
BB117D & 2000/02/05 &0.874 & -0.469 &-0.73 &8.1  & $61\pm5$  &70 &-     & -    &\tablenotemark{f}\\
BB117E & 2000/02/27 &0.881 & -0.505 &-0.76 &9.3  & $71\pm6$  &70 &-     & -    &\tablenotemark{f}\\
BB117F & 2000/03/19 &0.889 & -0.581 & 0.02 &7.4  & $103\pm15$&90 &-     & -    &\tablenotemark{f}\\
BB124A & 2000/04/20 &0.900 & -0.569 &-0.63 &6.6  & $70\pm8$  &76 &-     & -    &\tablenotemark{f}\\
BB124B & 2000/05/26 &0.912 & -0.591 &-1.01 &5.9  & $78\pm5$  &81 &-1.038 &-0.86 &8.7\\
BB124C & 2000/06/19 &0.920 & -0.543 & 0.14 &3.7  & $90\pm4$  &81 &-0.966 &1.93  &5.8\\
BB124D & 2000/07/21 &0.931 & -0.629 & 0.15 &3.9  & $80\pm7$  &79 &-1.012 &2.96  &7.2\\
BB125A & 2000/08/20 &0.942 & -0.668 & 0.41 &3.3  & $90\pm4$  &76 &-0.963 &2.07  &3.0\\
BB125B & 2000/09/25 &0.954 & -0.804 &-0.56 &1.9  & $96\pm5$  &83 &-     & -    & $<0.3$\tablenotemark{g}\\
BB125C & 2000/10/22 &0.963 & -      & -    &$<0.3$& -        & &-     & -    & $<0.3$\\
BB125D & 2000/11/18 &0.973 & -      & -    &$<0.2$& -        & &-     & -    & $<0.3$\\
\enddata
\tablenotetext{a}{Based on orbit period and epoch of periastron from \citet{Marchenko:2003}}
\tablenotetext{b}{Offset from $\alpha$=$20^h 20^m 27.977444^s$ $\delta$= $43\deg 51' 16.29174''$}
\tablenotetext{c}{Derived using {\sc aips} routine {\sc jmfit}}
\tablenotetext{d}{Position angle of the orientation of the emission arc - see \textsection{\ref{sec:vlba_obs}}}
\tablenotetext{e}{$\Theta$ is the observed opening angle of the WCR - see \textsection{\ref{sec:distance}}}
\tablenotetext{f}{Unable to adequately calibrate antenna phase - see \textsection{\ref{sec:vlba_obs}}}
\tablenotetext{g}{$3\sigma$ upper limit}
\end{deluxetable*}

The 8.4-GHz emission detected by the VLBA is clearly resolved.  We
identify this emission as arising from the WCR in WR\thinspace140
since this is the only emission in the system with sufficient
brightness temperature ($\sim10^{6-7}$~K) to be detected by the VLBA.
The stellar winds, with brightness of $\sim10^4$~K are undetected.  A
bow-shaped ridge of emission is observed at most epochs, as
anticipated for the WCR from model calculations
\citep[e.g. see][]{Eichler:1993, Canto:1996, Dougherty:2003a}, with
the bow shock wrapping around the star with the lower wind momentum -
typically the O star. A bow-shaped WCR has been suggested from
observations of other CWB systems \citep[e.g. see][]{Williams:1997,
Dougherty:2002}, but not as clearly as in the observations presented
here.

Between orbit phase 0.74 and 0.95, the WCR exhibits proper motion that
is largely from east to west, moving $\sim10$~mas over the observing
period, and also appears to rotate from ``pointing'' NW to W.  These
observations are key to unraveling the orbital motion of
WR\thinspace140 which we discuss more fully in
\textsection{\ref{sec:orbit}}.

It appears the WCR has a similar extent (and flux) from phase
0.737 to 0.841, and then decreases in size, notably after phase 0.874,
and again after phase 0.931.  We have estimated the brightness
temperature of the WCR through modelling of the visibilities at 8.4
GHz using the {\sc difmap} analysis package at orbital phases 0.737,
0.858 and 0.931. At phase 0.737, the WCR extends out to $\sim3.5$ mas
from its centre and has a brightness temperature of
$\sim3\times10^7$~K. This brightness temperature is essentially
constant at phases 0.858 and 0.931, even though the WCR is
considerably smaller by phase 0.931, only extending $\sim1.5$~mas from
the peak of emission.

Between phase 0.80 and 0.84, the arc of emission appears to develop
two peaks along its ridge, which then fade away and the arc is
reformed by phase 0.86. During these phases, it appears that the
intensity from the stagnation point is reduced. The concurrent VLA
observations (see \textsection{\ref{sec:vla_obs}}) show a marked variation
in flux at 1.5 GHz at this stage of the orbit. This emission is still
brightening from around 18 mJy at phase 0.80 to a peak of 26 mJy at
0.83, before fading to 20 mJy by phase 0.85. This is followed by the
suggestion of another slight brightening before the onset of the
decline to minimum. There is little evidence of any similar
brightening in the emission at the other observed frequencies.

\begin{deluxetable*}{cccccccc}
\tabletypesize{\scriptsize}
\tablecaption{VLA Observations of WR\thinspace140\label{tab:vla_obs}}
\tablewidth{0pt}
\tablehead{
\colhead{Obs. Date} & 
\colhead{Phase\tablenotemark{a}} & 
\colhead{1.5\GHz} &
\colhead{4.9\GHz } & 
\colhead{8.4\GHz} & 
\colhead{15\GHz} & 
\colhead{22\GHz} &
\colhead{Array} \\
&& (mJy) & (mJy) & (mJy) & (mJy) & (mJy) 
}
\startdata
 1998/06/28 & 0.671 &  $4.67\pm0.10$\tablenotemark{b} & $19.24\pm0.09$ & $25.14\pm0.08$ & $19.36\pm0.24$ & $17.91\pm0.39$ & BnA \\ 
 1999/05/19 & 0.783 & $16.94\pm1.90$ & $25.96\pm0.12$ & $22.80\pm0.07$ & $19.51\pm0.18$ &                & D \\ 
 1999/06/13 & 0.792 & $16.99\pm0.06$ & $25.94\pm0.06$ & $22.44\pm0.06$ & $18.78\pm0.17$ & $17.87\pm0.42$ & DnA \\ 
 1999/07/06 & 0.800 & $18.50\pm0.07$ & $25.94\pm0.06$ & $21.77\pm0.16$ & $20.32\pm0.22$ & $18.55\pm0.58$ & A \\ 
 1999/07/13 & 0.802 & $18.75\pm0.07$ & $26.50\pm0.07$ & $22.60\pm0.08$ & $18.77\pm0.18$ & $16.77\pm0.62$ & A \\ 
 1999/07/31 & 0.809 & $19.42\pm0.07$ & $25.04\pm0.09$ & $21.49\pm0.08$ & $18.61\pm0.19$ & $15.20\pm0.52$ & A \\ 
 1999/10/03 & 0.831 & $26.37\pm0.07$ & $27.40\pm0.07$ & $22.66\pm0.07$ & $18.40\pm0.19$ & $16.28\pm0.22$ & BnA \\ 
 1999/10/21 & 0.837 & $24.08\pm0.07$ & $24.52\pm0.07$ & $20.79\pm0.04$ & $16.85\pm0.20$ & $14.84\pm0.23$ & BnA \\ 
 1999/11/06 & 0.842 & $22.23\pm0.09$ & $24.73\pm0.08$ & $21.00\pm0.05$ & $16.17\pm0.16$ & $14.65\pm0.21$ & B \\ 
 1999/11/21 & 0.848 & $21.18\pm0.12$ & $23.48\pm0.09$ & $20.38\pm0.06$ & $17.18\pm0.21$ & $16.87\pm0.26$ & B \\ 
 1999/12/04 & 0.852 & $20.09\pm0.18$ & $21.87\pm0.09$ & $18.65\pm0.07$ & $16.77\pm0.29$ &                & B \\ 
 1999/12/27 & 0.860 & $22.54\pm0.20$ & $22.29\pm0.09$ & $18.68\pm0.12$ &                &                & B \\ 
 1999/12/29 & 0.861 & $21.58\pm0.17$ & $20.92\pm0.09$ & $17.80\pm0.09$ & $14.57\pm0.23$ & $13.46\pm0.21$ & B \\ 
 2000/01/14 & 0.866 & $20.54\pm0.12$ & $19.99\pm0.06$ & $16.74\pm0.06$ & $13.44\pm0.17$ & $12.59\pm0.18$ & B \\ 
 2000/02/23 & 0.880 & $24.00\pm0.26$ & $20.63\pm0.06$ & $17.55\pm0.07$ &                &                & CnB \\ 
 2000/03/29 & 0.892 & $22.22\pm0.39$ & $18.50\pm0.06$ & $15.76\pm0.07$ &                & $11.79\pm0.24$ & C \\ 
 2000/04/14 & 0.898 & $20.59\pm0.57$ & $17.51\pm0.06$ & $15.22\pm0.06$ &                & $11.68\pm0.21$ & C \\ 
 2000/05/11 & 0.907 & $17.29\pm0.28$ & $15.27\pm0.07$ & $13.52\pm0.06$ & $11.71\pm0.20$ & $ 9.52\pm0.16$ & C \\ 
 2000/06/21 & 0.921 & $13.56\pm0.44$ & $13.59\pm0.07$ & $12.39\pm0.09$ & $11.81\pm0.20$ & $10.00\pm0.38$ & DnC \\ 
 2000/07/17 & 0.930 & $ 8.60\pm0.57$ & $10.00\pm0.07$ & $ 9.71\pm0.07$ & $ 9.95\pm0.11$ & $10.19\pm0.23$ & DnC \\ 
 2000/09/27 & 0.955 & $ 2.35\pm0.74$ & $ 4.74\pm0.12$ & $ 6.24\pm0.06$ & $ 6.94\pm0.09$ & $ 7.48\pm0.14$ & D \\ 
 2000/10/15 & 0.961 & $ 1.20\pm0.09$ & $ 3.51\pm0.07$ & $ 4.48\pm0.07$ &                & $ 7.16\pm0.25$ & A \\ 
 2000/11/21 & 0.974 & $ 0.62\pm0.07$ & $ 1.66\pm0.05$ & $ 2.64\pm0.06$ &                & $ 5.00\pm0.11$ & A \\ 
\enddata
\tablenotetext{a}{Based on orbit period and epoch of periastron from \citet{Marchenko:2003}}
\tablenotetext{b}{1\,$\sigma$ errors measured in outlying image regions.}
\end{deluxetable*}

A similar two-peaked structure was also apparent in MERLIN
observations of WR\thinspace147, a very wide WR+OB CWB
\citep{Watson:2002}. It is difficult to attribute the reduced level of
emission between the peaks seen in WR\thinspace147 to higher free-free
extinction along that line-of-sight relative to the peaks because
the line-of-sight extinction is very low in such
a wide binary system. Alternatively, such a twin-peak structure may be
the result of time-variable wind density resulting in changes in the
post-shock density in the WCR.  Certainly, between phase 0.80 and 0.84
the 15-GHz VLA observations of WR\thinspace140 (see
Fig.~\ref{fig:light_curve}) have large scatter that may be related to
such structure changes. We draw attention to a similar structure seen
in the synthetic 5~GHz image of WR\thinspace147 produced by
\citet[][Figure 17]{Dougherty:2003a}. However, in this case, the
structure is attributed to numerical instabilities in the
hydrodynamics code used to generate the model (J. Pittard,
priv. communication).

By phase 0.963, the WCR is no longer detected by the VLBA. This was
expected from the radio light curve (see Fig.~\ref{fig:light_curve})
since by this stage of the orbit, the radio emission at all
frequencies has declined to less than a few mJy. Based on our
experience from these observations, we expect to be able to detect the
WCR again somewhere around phase 0.3-0.4 during the next orbit.

The WCR is also resolved in the 1.7-GHz observations (see
Fig.~\ref{fig:lband_vlba}). As at 8.4~GHz, proper motion of the source
is suggested as the orbit progresses.  However, the proper motion at
1.7~GHz does not appear as continuous as at 8.4~GHz.  Our difficulties
with the 1.7-GHz data can be attributed to rapidly varying
differential ionospheric delay and phase offsets between the position
of the phase-reference source and WR\thinspace140 during the
observations. Throughout the observing campaign, solar activity was
close to the maximum of its 11-year cycle, contributing to significant
and rapid ionospheric variability.

A summary of the positions and fluxes of the VLBA data is given in
Table~\ref{tab:vlba_obs}. The fluxes and positions were derived at
both 8.4 and 1.7~GHz using the {\sc aips} routine {\sc jmfit}. As a
check, fluxes were also determined using {\sc tvstat}, where values
within $\pm5$\% of those derived with {\sc jmfit} were found. This
difference is approximately the uncertainty that is to be expected in
the absolute flux scale calibration. However, no flux uncertainty is
quoted in Table~\ref{tab:vlba_obs} since the VLBA resolves out
$\sim40$\% of the emission from WR\thinspace140 (see
Sec.~\ref{sec:vla_obs}). The uncertainty in the positions is most
likely dominated by ionospheric delay and phase offset error,
certainly at 1.7~GHz. Since delay $\propto \nu^{-2}$ and offset
$\propto \nu^{-1}$, they are also likely to dominate at 8.4~GHz over
the Gaussian source fitting errors, which are a fraction of a
milli-arcsecond in these images. For epochs where the WCR emission
appears double-peaked, the mid-point of the two peaks is quoted.  Also
included in Table~\ref{tab:vlba_obs} are our estimates of the position
angle of the axis of symmetry of the WCR, along with their
uncertainty. We also include estimates of the opening angle of the
WCR. Though not entirely robust, we believe each of these estimates
are good to first order.

\subsection{VLA observations}
\label{sec:vla_obs}
\begin{figure}
\includegraphics[angle=0,scale=0.7]{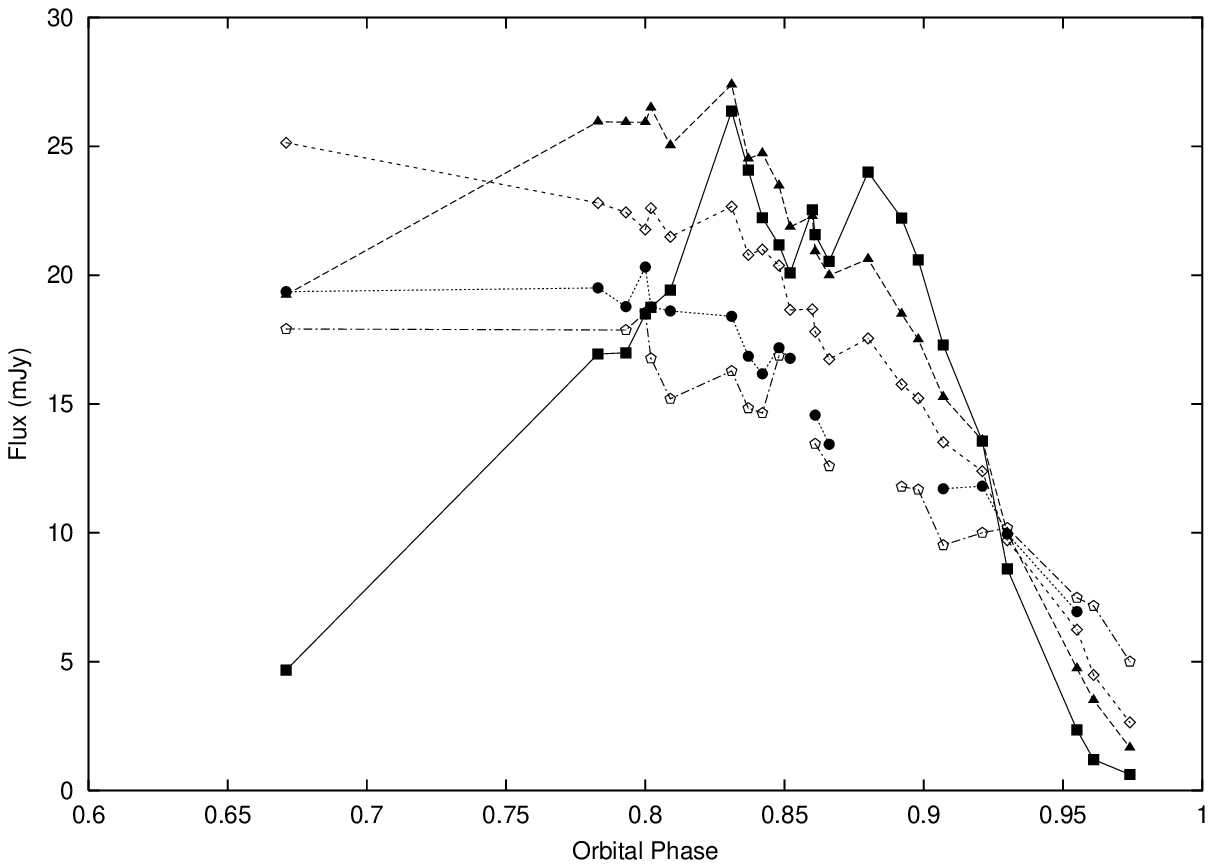}
\caption[]{The VLA observations of WR\thinspace140 from
Table~\ref{tab:vla_obs} as a function of orbital phase at 1.5
(squares), 4.9 (triangles), 8.4 (open diamonds), 15.0 (circles) and 22
GHz (open pentagons). The data points are connected to help guide the
eye. Gaps are where data were not collected at 15 and 22~GHz.
\label{fig:new_vla_data}}
\end{figure}

Multi-frequency VLA observations of WR\thinspace140 were made
throughout the VLBA observations to provide continuum spectral
information.  Observations were made in standard continuum mode, using
two 50-MHz channels centered on 1465 (20cm), 4885 (6cm), 8435 (3.6cm),
14965 (2cm), and 22485 MHz (1.3cm). Antenna phases were established
via phase-referencing using J2007+404, and the absolute flux level
from observation of the bright source 3C286. Data was calibrated and
reduced in the standard manner using the {\sc aips} package. Since
WR\thinspace140 is unresolved in these observations, fluxes were
determined using {\sc jmfit} assuming a point source. The results of
the VLA observations are given in Table~\ref{tab:vla_obs} and
displayed in Fig.~\ref{fig:new_vla_data}.  The quoted uncertainties
are the background rms of the images. We note that the absolute flux
scale has an uncertainty of $\sim5$\%.

\begin{figure}
%\hrule
\vspace{18.6cm}
%\hrule
\includegraphics{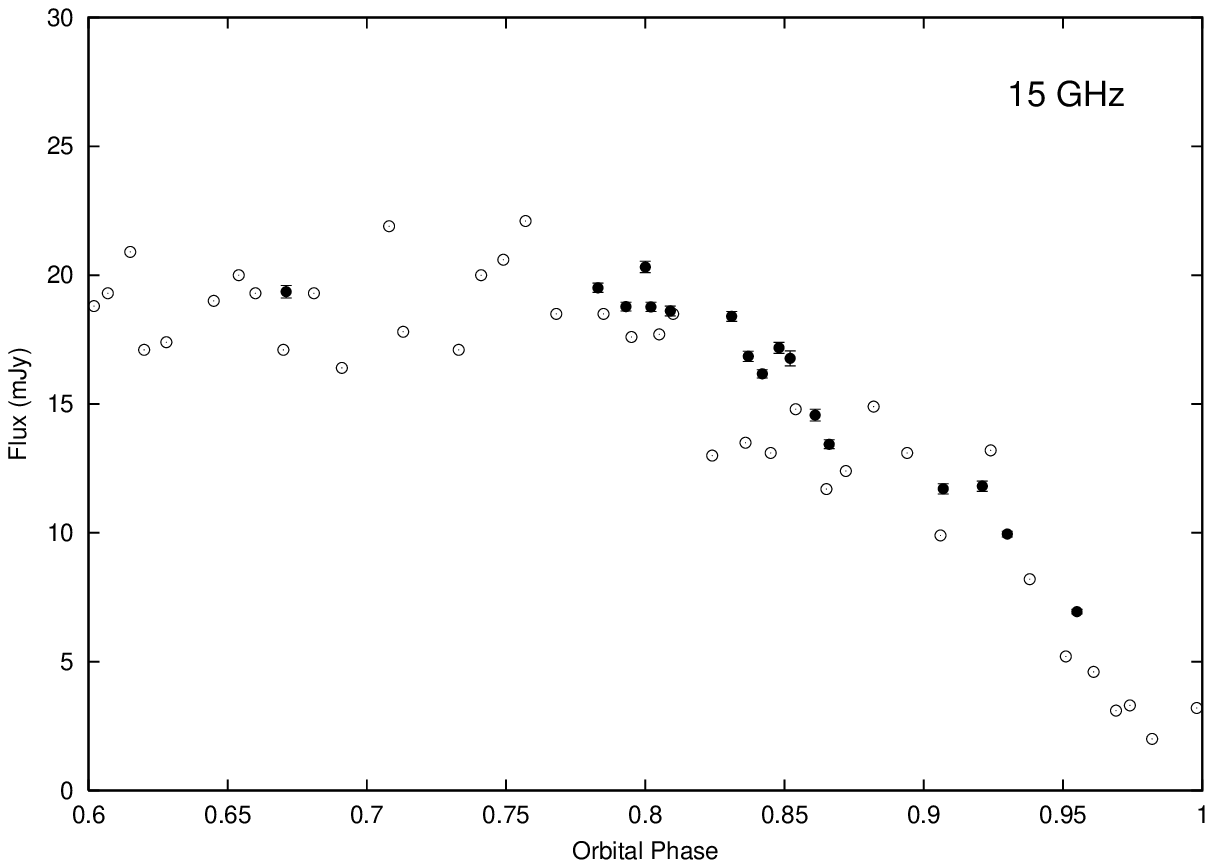}
\includegraphics{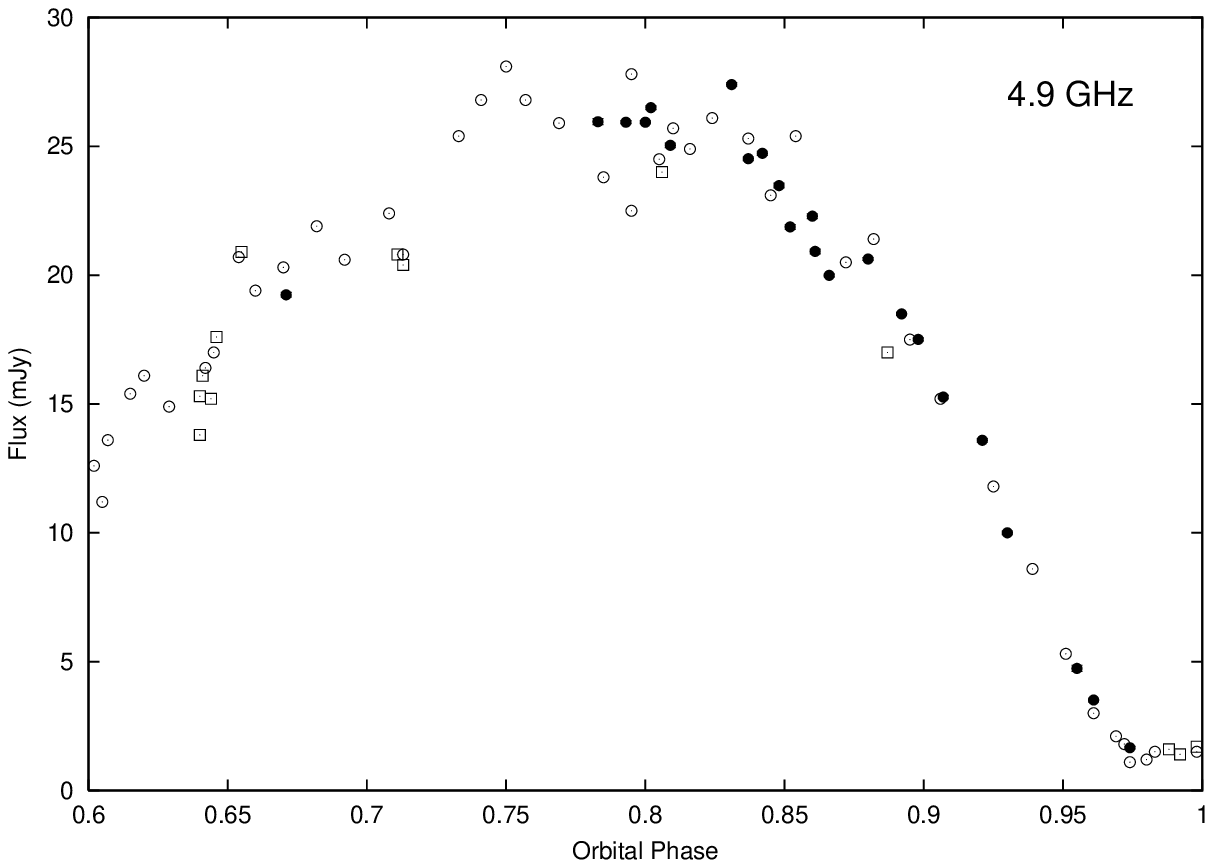}
\includegraphics{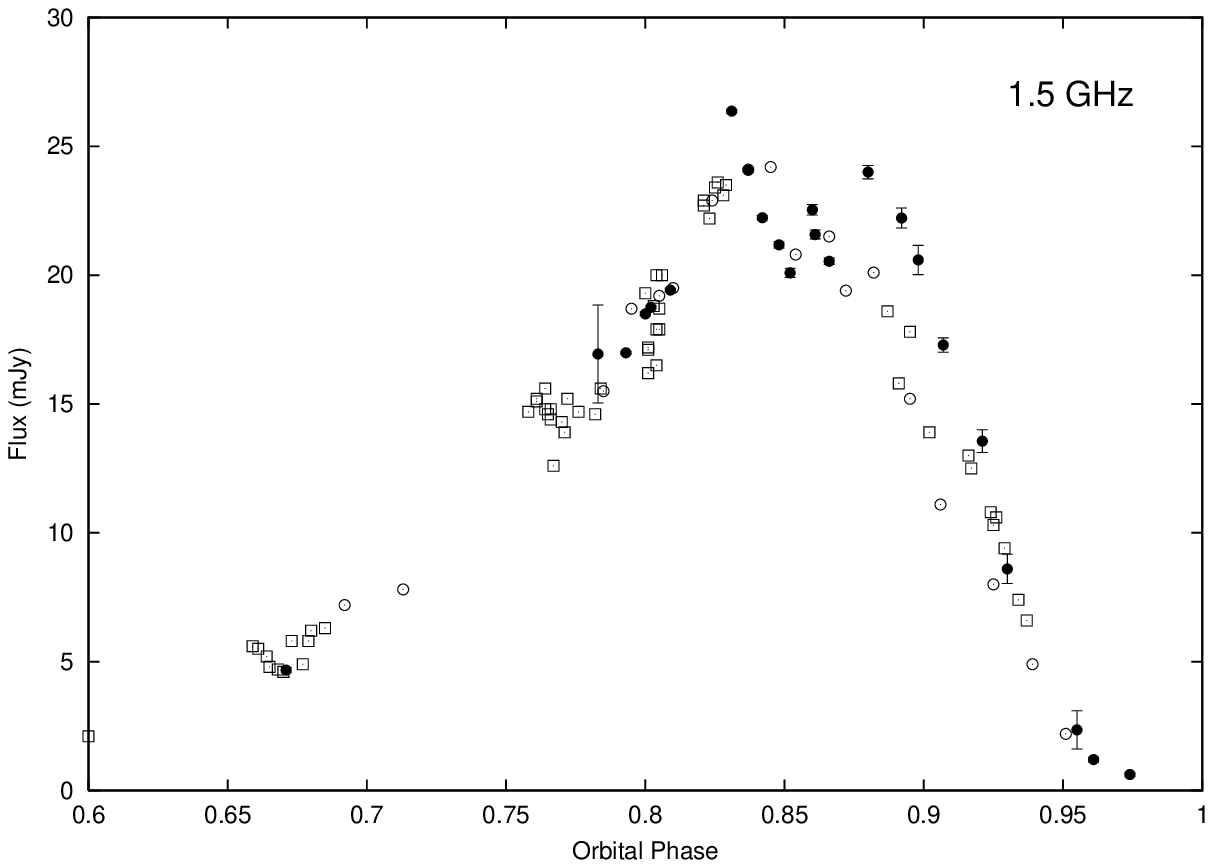}
\caption[]{The new VLA observations (solid circles) compared with
observations from previous orbits at 15 (top), 4.9 (middle) and 1.5
GHz (bottom). Open circles are previous VLA observations
\citep{White:1995} and open squares are from the WSRT
\citep{Williams:1990, Williams:1994}
\label{fig:new_vs_old}}
\end{figure}

The VLBA fluxes at 8.4~GHz are closely 50\% of the VLA 8.4-GHz
fluxes.  Assuming the thermal flux at 8.4 GHz is that at phase 0.974
(2.64~mJy), we estimate the VLBA is resolving out $\sim40$\% of the
synchrotron flux. This missing flux is most likely emission
arising further from the shock apex than the emission detected 
by the VLBA.  At 1.7~GHz, the VLBA fluxes have large
scatter, most certainly due to calibration problems. As a result, the
relationship to the VLA observations is unclear. It appears that
40-50\% of the emission is undetected at phases around 0.8, though
there could be evolution of this fraction with orbital phase. At phase
0.75 and near phase 0.95, there may be little or no flux resolved out,
suggesting that at these phases the source may be smaller than between
these phases.

In Fig.~\ref{fig:new_vs_old}, the new data are directly compared with
those measured by others from the orbit immediately prior to the one
presented here.  The new data shows less scatter than the earlier
observations, which we attribute to improved system performance and
better antenna phase calibration. It is clear from these figures that
the radio emission from WR\thinspace140 very closely repeats from one
orbit to the other. The radio light curves at each of the frequencies
are very similar, with the peak and the decline to minimum also being
very similar.  There may be an offset between the 1.5~GHz data from
phase 0.88 to 0.92. However, we caution against over interpreting this
apparent difference.  The scatter in the previous 1.5~GHz fluxes
between orbital phase 0.75 and 0.82 leads us to think that the
apparent offset may not be significant.  The close similarity between
the fluxes during successive orbits at all three frequencies leads us
to conclude that whatever process(es) govern the nature of the radio
emission from WR\thinspace140, they repeat from one orbit to another,
suggesting that they are controlled primarily by orbital motion.

\subsection{MERLIN observations}
\label{sec:merlin_obs}
In conjunction with the VLBA observations, observations using the
Multi-Element Radio-Linked Interferometer Network (MERLIN) in the UK
were obtained in December 1999 and February 2000 at 1.66~GHz. The
observations were approximately 10 hours duration, with a total
bandwidth of 16 MHz in each of two polarizations. In addition,
observations at 5~GHz on June 1992 and 1.67~GHz on August 1993
were extracted from the MERLIN archive.  All of these observations
were phase referenced using J2007+404, and reduced in the standard
manner with the {\sc aips} program. The absolute flux scale was
determined through observations of 3C286 and bootstrapped to the
bright point-sources OQ208 and 0552+298. WR\thinspace140 is not
resolved in the MERLIN observations at either 1.6 or 5~GHz. The
salient results of these observations are summarised in
Table~\ref{tab:merlin_obs} where the quoted fluxes are determined
assuming a point source, and taking the rms background level of the
images to be the flux uncertainty. As in the VLA observations, the
absolute flux scale has an uncertainty of $\sim5$\%.

\begin{deluxetable}{cccc}
\tabletypesize{\scriptsize}
\tablecaption{MERLIN Observations of WR\thinspace140\label{tab:merlin_obs}}
\tablewidth{0pt}
\tablehead{
\colhead{Obs. Date} & 
\colhead{Phase\tablenotemark{a}} & 
\colhead{1.67\GHz} &
\colhead{5\GHz }  \\
&& (mJy) & (mJy)
}
\startdata
 1992/06/28 & 0.906 &  &$13.2\pm0.2$ \\
 1993/08/06 & 0.055 & $<0.3$ &  \\
 1999/12/29 & 0.861 & $20.3\pm0.1$ &  \\
 2000/02/02 & 0.873 & $24.4\pm0.2$ &  \\
 2000/02/05 & 0.874 & $24.2\pm0.1$ &  
\enddata
\tablenotetext{a}{Based on orbit parameters from \citet{Marchenko:2003}}
\end{deluxetable}

Comparison of the MERLIN fluxes and the VLA observations show very
similar flux levels, as expected for a source that is unresolved by
both of these interferometer arrays. %(Fig.~\ref{fig:flux_compare})
Since WR\thinspace140 is unresolved by MERLIN, we estimate the extent
of the emission region is less than 13~mas at 5~GHz.

\section{The orbit and distance of WR140}
\begin{figure}
\includegraphics[angle=0,scale=0.69]{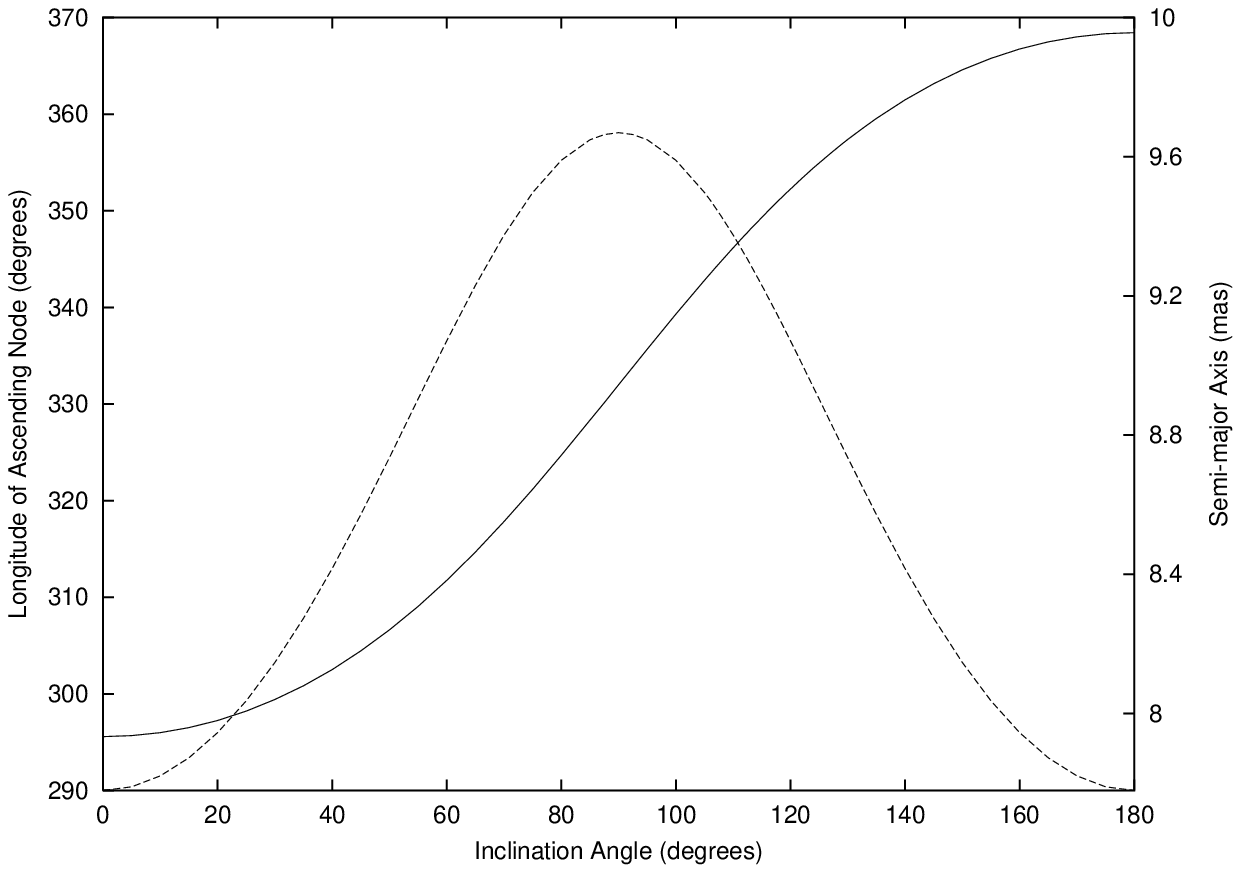}
\caption[]{Solutions for the longitude of the ascending node (solid
line) and orbit semi-major axis (dotted line) as a function of orbit
inclination, derived from an observed separation of $12.9$~mas at a
position angle of $151.7\deg$ on June 17, 2003 (orbit phase 0.297)
\citep{Monnier:2004}.The uncertainty in the IOTA position angle gives
a spread in the possible values of $\Omega$ of closely $\pm1\deg$,
whereas the uncertainty in the separation gives $\pm0.3$~mas for the
semi-major axis uncertainty.\label{fig:solution_family}}
\end{figure}

\subsection{Orbital parameters of WR\thinspace140}
\label{sec:orbit}
Many of the orbital parameters in WR\thinspace140, in particular the
orbital period ($P$), epoch of periastron passage ($T_o$),
eccentricity ($e$) and the argument of periastron ($\omega$) are
well-determined from radial velocity measurements \citep[see][and
references therein]{Marchenko:2003}. However, the orbital inclination
($i$), semi-major axis ($a$) and the longitude of the ascending node
($\Omega$) require that the system is resolved into a ``visual''
binary. Recently, the two stellar components in WR\thinspace140 have
been resolved using the Infrared-Optical Telescope Array (IOTA)
interferometer at a single epoch \citep{Monnier:2004}. This single
observation allows for the first time the determination of families of
solutions for $(i, a, \Omega)$. However, until further IOTA
observations are available, the VLBA observations of the WCR offer the
only means to determine the orbit direction and inclination, and
hence $\Omega$ and $a$.

\citet{Monnier:2004} observed WR\thinspace140 on June 17, 2003 to have
a separation of $12.9^{+0.5}_{-0.4}$~mas at a position angle of
${151.7^{+1.8}_{-1.3}}$~degrees east of north.  Assuming
$P=2899$~days, $T_o=2446147.4$, $e=0.881$ and $\omega=47\deg$
\citep{Marchenko:2003}, this observation at orbital phase 0.297 gives
families of solutions for ($i,\Omega,a$) as shown in
Fig.~\ref{fig:solution_family} for $0\deg<i<180\deg$. We examine
the full range of possible inclination angles since the direction of
relative orbital motion dictates whether $i$ is greater or smaller
than $90\deg$. 

Low orbit inclinations angles $i<30\deg$ (or equivalently $i>150\deg$)
can be easily ruled out since radial velocity variations are observed
and low inclinations would require the stellar components to have
untenably large masses. Ultra-violet line observations with IUE,
thought to have been obtained close to conjunction estimate
$i\sim38\deg$ \citep{SetiaGunawan:2001}. The same analysis with
orbital elements from \citet{Marchenko:2003} give higher estimates of
$i$, with larger uncertainty (P.M. Williams, priv. communication).
\citet{Marchenko:2003} estimated $50\pm15\deg$ from their optical line
study, and similar work by \citet{Varricatt:2004} using the He~{\sc i}
emission line in the near-IR gives an estimate of $i\simeq 65\deg$.
Unfortunately, these estimates are difficult to constrain well, and
all have large uncertainty.

\begin{figure}
\includegraphics[angle=0,scale=0.7]{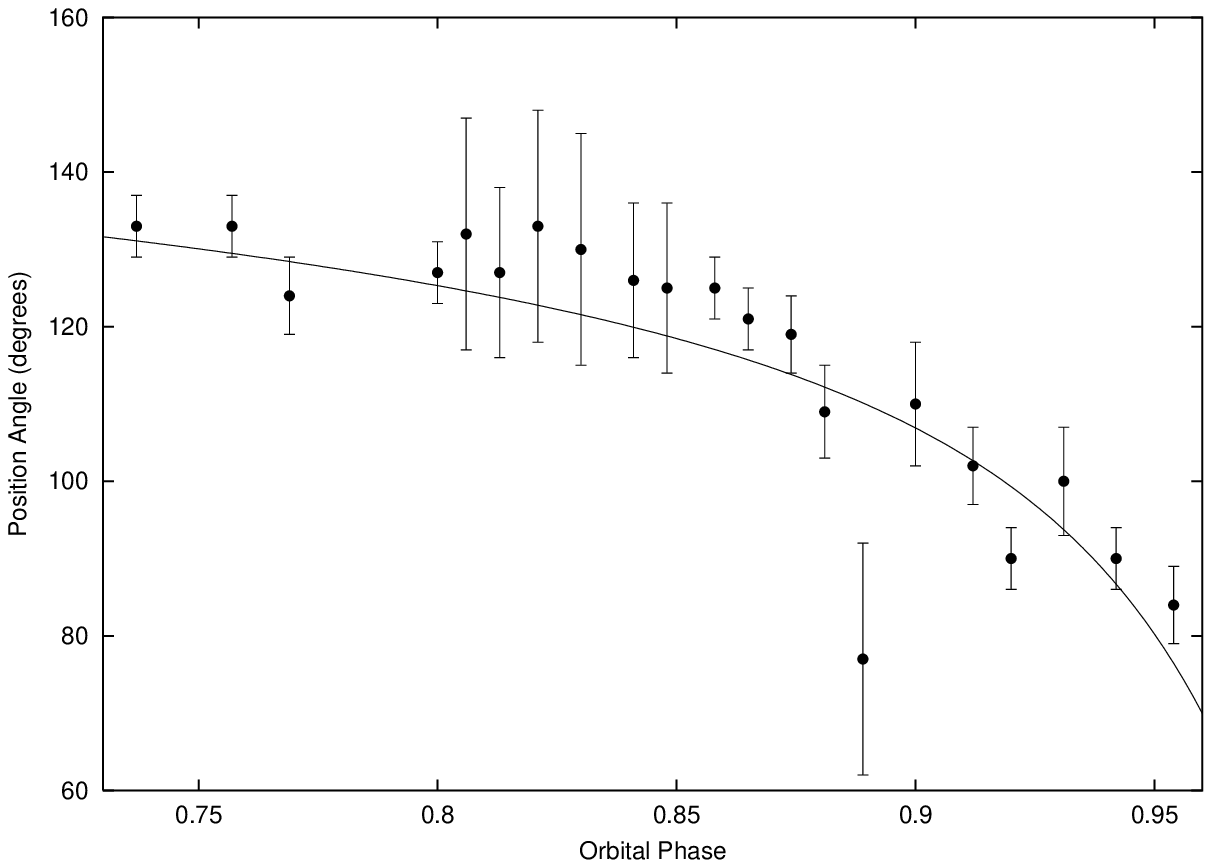}
\caption[]{The change in the position angle of the axis of symmetry of
the WCR as a function of orbital phase, as estimated from the VLBA
8.4-GHz observations (see Table~\ref{tab:vlba_obs}).  The solid line
is the weighted best-fit curve of the position angle of the
line-of-centres of the two stellar components projected on the plane
of the sky as a function of orbital phase, deduced using the orbit
parameters from \citet{Marchenko:2003} and \citet{Monnier:2004}. The
line corresponds to $i=122\deg$ and $\Omega=353\deg$.
\label{fig:posinc}}
\end{figure}

The VLBA observations of the WCR permit us to constrain acceptable
solutions for $i$. Under the assumption that the free-free opacity
along the line-of-sight to the WCR is sufficiently low as to not
impact the apparent distribution of emission from the WCR, we expect
the arc of WCR emission to wrap around the star with the lower wind
momentum (the O star) and ``point'' toward the WR star.  In this case,
the rotation of the orientation of the WCR as the orbit progresses
implies that the O star moves from SE to close to due E of the WR star
over the period of the VLBA observations. If it is assumed the axis of
symmetry of the WCR emission is coincident with the projection on the
plane of the sky of the line-of-centres of the two stars in the
binary, then we can derive the orbital inclination from the change in
the orientation of the WCR with orbital phase.  Each ($i,\Omega$)
family provides a unique set of position angles for the projected
line-of-centres as a function of orbital phase. Using a weighted
minimum $\chi^2$ measure to fit the position angle of the line of
symmetry of the WCR as a function of orbit phase for different sets of
($i,\Omega$), we find a best-fit solution of $i=122\deg\pm5$ and
$\Omega=353\pm3$ (Fig.~\ref{fig:posinc}).  These values lead to a
value for the semi-major axis of $a=9.0\pm0.5$~mas, where the
uncertainty is dominated by the uncertainty in the separation as
observed by IOTA.

\begin{figure}
\vspace{17.3cm}
\includegraphics{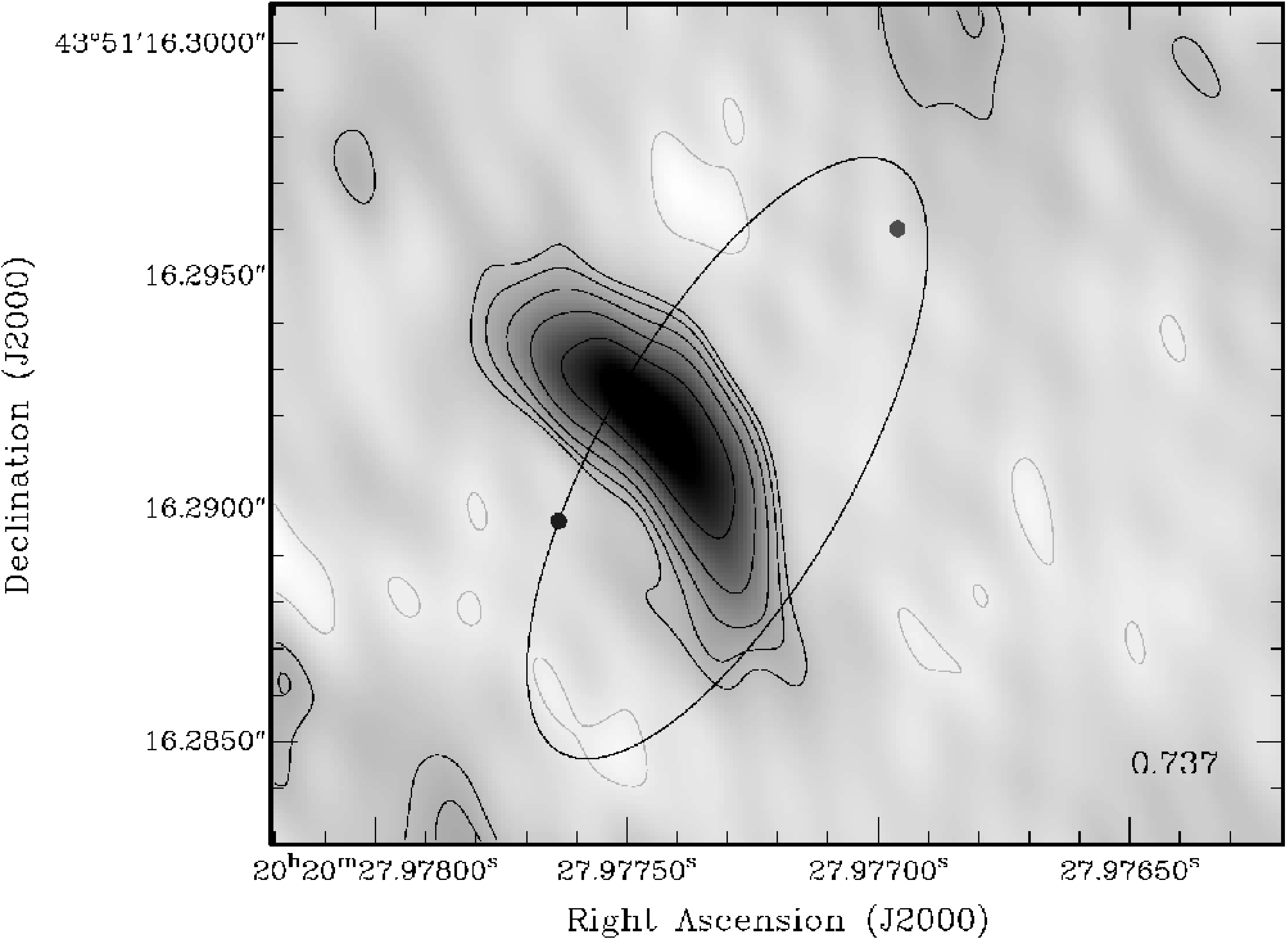}
\includegraphics{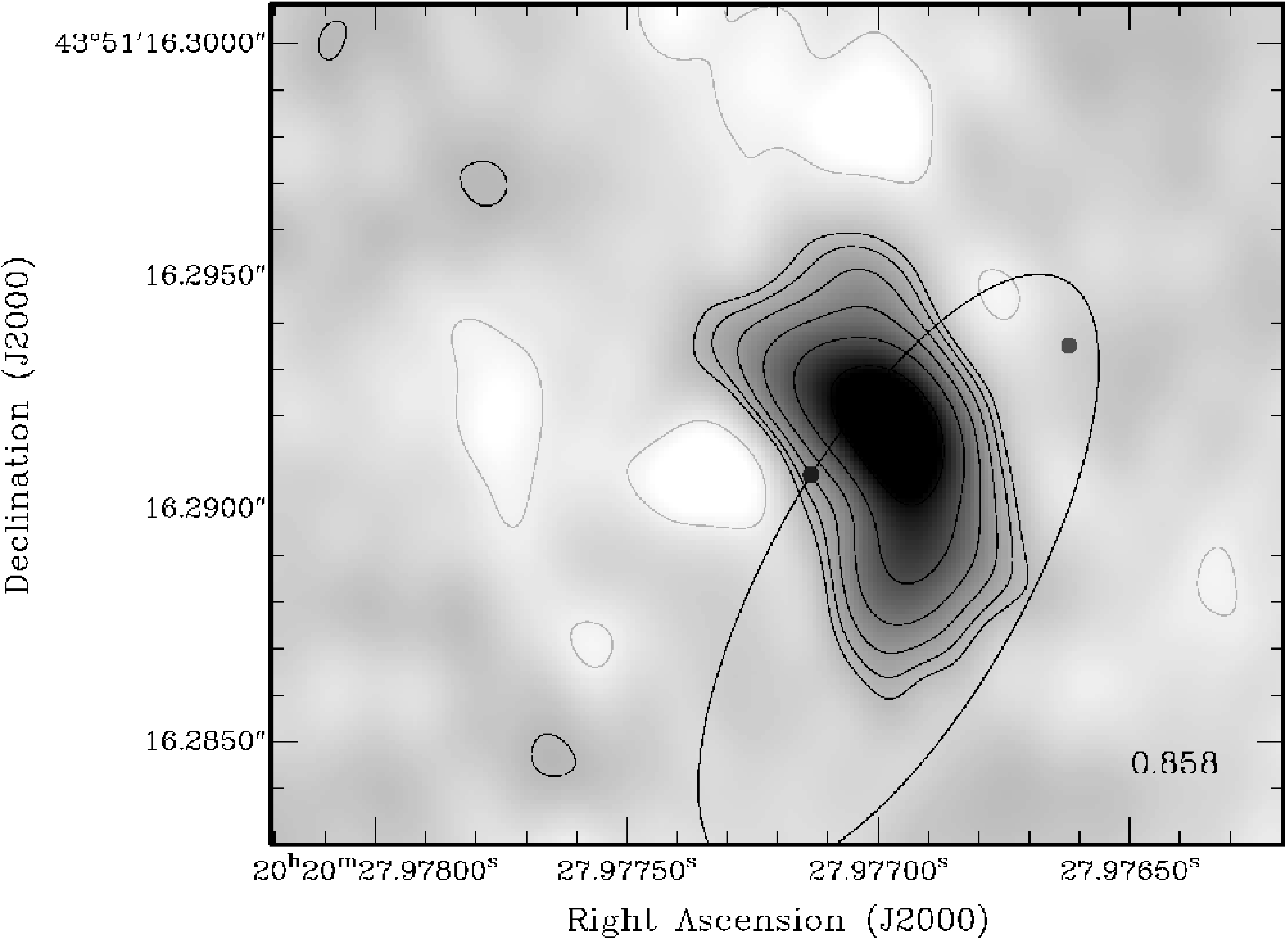}
\includegraphics{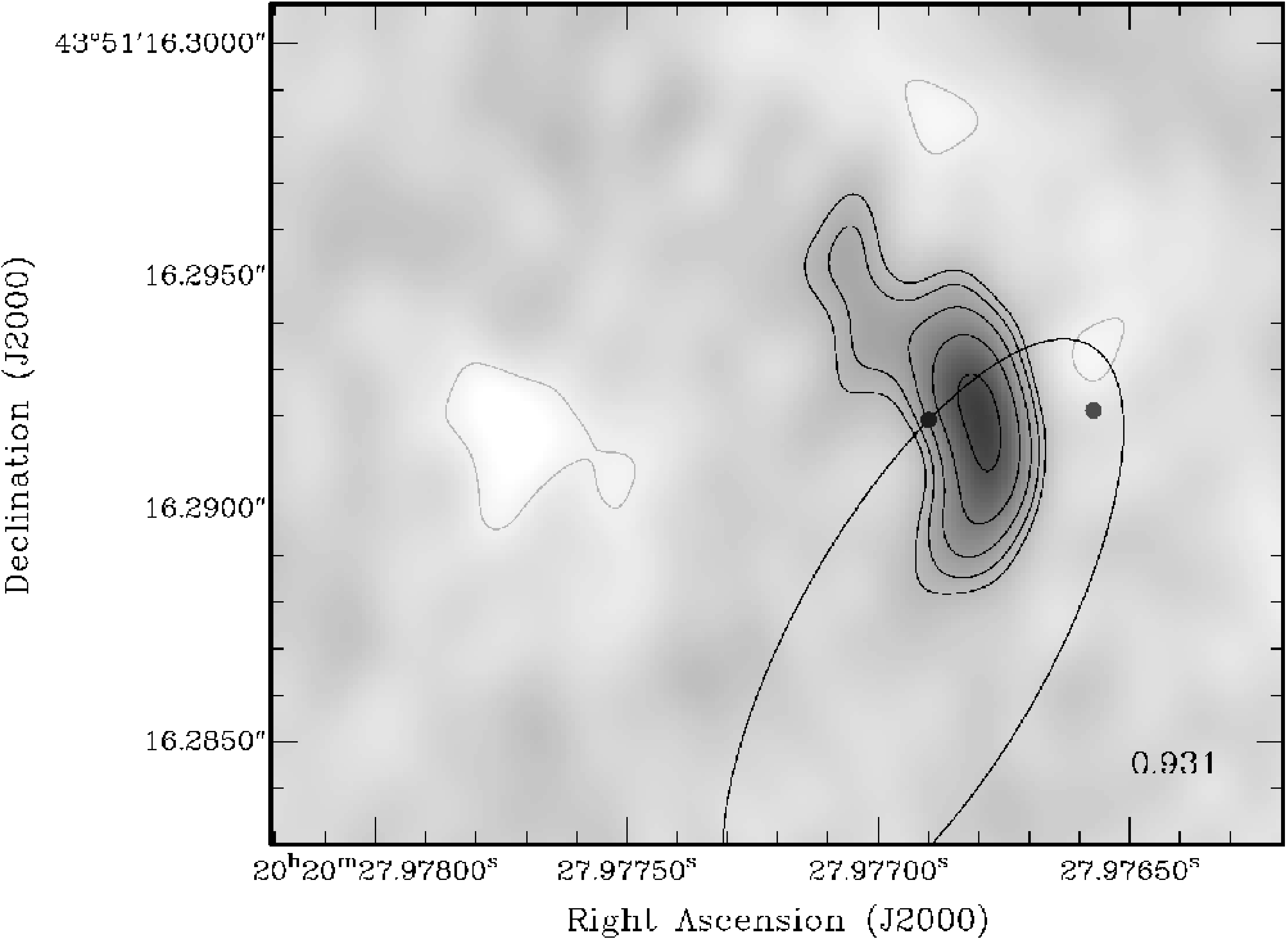}
\caption[]{The orbit of WR\thinspace140 on the plane of the sky using
$e=0.88$, $\omega=47\deg$, $\Omega=353\deg$, $i=122\deg$ and
$a=9.0$~mas at orbital phase 0.737, 0.858 and 0.931, overlaid on the
VLBA 8.4-GHz images. The WR star is to the W (right) of the WCR at
these phases.  The proper motion and the rotation of the WCR as the
orbit progresses is clear. The relative position of the stars and the WCR
was determined using a wind momentum ratio of $0.22$ (see
\textsection{\ref{sec:distance}}). 
\label{fig:orbit_vlba}}
\end{figure}

Using the semi-major amplitudes of the radial velocity curve,
eccentricity and period from \citet{Marchenko:2003}, an orbit
inclination of $i=122\deg\pm5\deg$ gives masses of
$20\pm4$~M$_\odot$ and $54\pm10$~M$_\odot$ for the WC and O star
respectively, within the anticipated mass ranges of a WC7 and an O4-5
star. We note that the uncertainty in the masses is dominated by that
in the orbital inclination.

The derived orbit inclination is consistent with values previously
suggested. However, this newly derived orbit presents a challenge to
current models of dust formation. To date, most models of dust
formation in WR\thinspace140 assume that the gas from which dust is
formed in the WCR, is compressed within $\sim0.15$~yr of periastron
passage, when densities in the WCR are sufficiently high for dust
formation to occur \citep{Williams:1990}. The
subsequent motion of the compressed gas is determined by the velocity
of this material when it is compressed.  Since the momentum of the WR
star wind is higher than that of the O star, this material moves away
from the WR star, along the WCR.  With the orbit orientation derived
here, the O star is NW of the WR star during periastron, and material
compressed at periastron will therefore have a proper motion to the
NW. However, this is not what is observed. High-resolution IR
observations show that dust ejected during the 2001 periastron passage
has proper motion from west to east, away from the WR star
\citep{Monnier:2002}. New dust models are now attempting to address
this challenge \citetext{Williams 2005, in preparation}.

If the WR star was to NW of the O star at periastron, as previously
hypothesized from the dust proper motion, the WCR would then have to
be wrapped around the WR star in the VLBA observations. This requires
the unlikely scenario of the O-star wind momentum being greater than
that of the WR star wind.  We acknowledge that free-free opacity of
the stellar winds can effect the observed shape of the WCR, but we
anticipate the free-free opacity in WR\thinspace140 would only be
sufficiently large to have a major impact on the apparent distribution
of the 8.4-GHz emission close to periastron \citep[see][Fig.11 for an
example]{Dougherty:2003a}. However, until full hydrodynamical
simulations of WR\thinspace140 are available, the full impact of
free-free opacity on the appearance of the WCR remains unclear. High
resolution near-IR observations at several bands should be able to
distinguish the identity of the two stars since we anticipate the WR
star is significantly brighter at longer IR wavelengths than the O
star.

\begin{deluxetable}{llll}
\label{tab:wr140_parms}
\tabletypesize{\scriptsize}
\tablecaption{Basic parameters of WR\thinspace140\label{tab:stellar_parms}}
\tablewidth{0pt}
\tablehead{\colhead{Parameter} & \colhead{Primary} & \colhead{Secondary}& \colhead{System} }
\startdata
Distance (kpc) & & & 1.85 \\
M$_v$ & -6.4 & -5.6 & -6.8\\
Spectral type & O4-5 I \tablenotemark{a} & WC7 & \\
BC    & -4.3 \tablenotemark{a} & -3.4 \tablenotemark{b} & \\
M$_{\rm bol}$ & -10.7 & -9.0 & \\
log (L$_{\rm bol}$/L$_\odot$) & 6.18 \tablenotemark{c} & 5.5 \tablenotemark{c} & \\
Mass (M$_\odot$) & $54\pm10$ & $20\pm4$ \\
v$_\infty$ (km~s${-1}$) & 3100 \tablenotemark{d} & 2860 \tablenotemark{e}\\
$\dot{\rm M}$ (${\rm M}_\odot$~yr$^{-1}$ & $8.7\times10^{-6}$ & $4.3\times10^{-5}$& 
\enddata
\tablenotetext{a}{Based on M$_v$ and the calibration of \citet{Vacca:1996}}
\tablenotetext{b}{\citet{Williams:1990}}
\tablenotetext{c}{Calculated assuming M$_{\odot, bol}=4.75^m$ \citep{Allen:1976}}
\tablenotetext{d}{\citet{SetiaGunawan:2001}}
\tablenotetext{e}{\citet{Eenens:1994}}
\end{deluxetable}

\subsection{Distance of WR\thinspace140}
\label{sec:distance}
Distance estimates of WR stars are typically based on absolute
magnitude calibrations that often have large scatter
\citep[see][]{vanderhucht:2001}.  Having determined the orbital
inclination and semi-major axis it is now possible to make an estimate
of the distance to WR\thinspace140 {\em independent of any stellar
parameters}. \citet{Marchenko:2003} determined $a\,\sin i =
14.10\pm0.54$~AU from radial velocity observations, which leads to
$a=16.6\pm1.1$~AU for $i=122\deg\pm5\deg$. We have derived the
semi-major axis to be $a=9.0\pm0.5$~mas, and together these
give a distance of $1.85\pm0.16$~kpc.

\subsection{Basic system parameters of WR\thinspace140}
This distance is somewhat larger than the usually quoted value of
1.3~kpc deduced by \citet{Williams:1990} from the luminosity of the
system. Since the primary O-star luminosity indicator is masked by the
WC7 spectrum, \citet{Williams:1990} assumed a main sequence O4-5 star
with an absolute magnitude of $-5.6$ and took that of the WC7 star to
be $-4.8$. However, they remarked the distance estimate would increase
if the O star was more luminous than main sequence. With the system at
$1.85$~kpc, the absolute magnitudes of the O4-5 star and the WC7 star
become $-6.4$ and $-5.6$ respectively. An absolute magnitude of $-5.6$
for the WC7 star is bright compared to the mean values (between $-4.5$
and $-4.9$) deduced from previous estimates for WC7 stars
\citep[see][Table 27]{vanderhucht:2001} but the dispersion is large
($\sim1^{\rm m}$), and certainly the value presented here is not
extreme. The brightness of the O4-5 star suggests that it is a
supergiant \citep[see][Table 7] {Vacca:1996}.

With the increase in distance, a reassessment of the mass-loss rates
of the two stars is appropriate. To circumvent any uncertainties
presented by clumping corrections that are required when using radio
flux determinations of mass loss, we turn to X-ray observations in
order to try and determine the mass-loss rate of the WR star. The
X-ray luminosity in WR\thinspace140 is dominated by emission from the
shocked WR wind material \citep{Pittard:2002}, where any clumps are
assumed to have been destroyed by the shock. From \citet{Zhekov:2000},
the X-ray luminosity measured by ASCA gives a mass-loss rate for the
WR star at 1.85 kpc of $4.3\times10^{-5}~{\rm
M}_\odot$~yr$^{-1}$. This can be readily reconciled with the unclumped
estimate of $5.3\times10^{-5}~{\rm M}_\odot$~yr$^{-1}$ from
\citet{Williams:1990} with a reasonable filling factor. Since
mass-loss rates derived from thermal free-free emission $\propto
D^{1.5}$, the unclumped value of \citet{Williams:1990} for
WR\thinspace140 at 1.85~kpc becomes $9.7\times10^{-5}~{\rm
M}_\odot$~yr$^{-1}$. Comparison with the clumping-free, X-ray derived
value implies a filling factor $\sim0.2$. For the O star,
\citet{Repolust:2004} suggests values of $8.6-8.8\times10^{-6}~{\rm
M}_\odot$~yr$^{-1}$ for O4-5 supergiants.

\begin{figure}
\includegraphics[angle=0,scale=0.7]{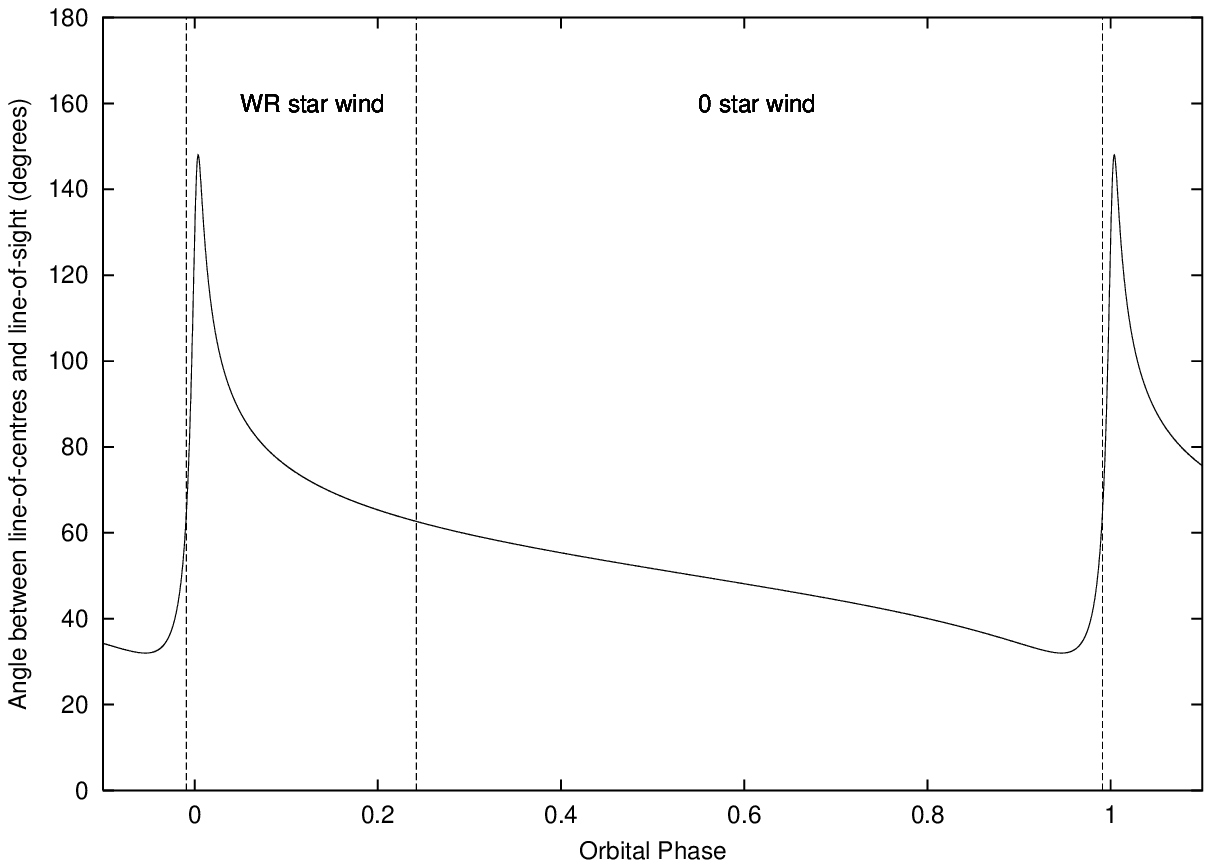}
\caption[]{The angle between the line-of-sight and the line-of-centres
of the two stars in WR\thinspace140 as a function of orbital phase.
The vertical lines at phases 0.24 and 0.99 are when this angle
transits through $63\deg$, the half-opening angle of the WCR shock
cone as deduced from $\eta=0.22$ and following \citet{Eichler:1993}.
\label{fig:losvsloc}}
\end{figure}

These mass-loss rates imply a wind momentum ratio $\eta=0.22$,
allowing us to position the stars relative to the WCR in the VLBA
images (see Fig.~\ref{fig:orbit_vlba}).  This wind-momentum ratio is
considerably higher than the 0.035 deduced by \citet{Williams:1990},
assuming a main sequence O star, and a mass-loss rate for the WR star
that was uncorrected for clumping. Following \citet{Eichler:1993},
$\eta=0.22$ implies a half-opening angle for the WCR of
$63\deg$. Using the VLBA images in Fig.~\ref{fig:xband_vlba} we can
estimate the half-opening angle $\theta$ from $$\tan \theta = \sin
\psi \tan(\Theta/2),$$ where $\psi$ is the angle between the
line-of-sight and the line-of-centres between the two stars, and
$\Theta$ is the observed full opening angle of the WCR. Using the
orbit derived in \textsection{\ref{sec:orbit}} to derive $\psi$ at the
observed orbital phases (Fig.~\ref{fig:losvsloc}), and the values of
$\Theta$ given in Table~\ref{tab:vlba_obs} we find
$\theta=65\deg\pm10\deg$, consistent with the estimate derived from
the wind-momentum ratio. Adding further support to this value,
\citet{Varricatt:2004} would derive a similar opening angle from
interpretation of He I line observations using the orbital inclination
derived here.

With the orbit size and wind-momentum ratio we can compare the
observed size of the emission region with $\pi r_O$, the size of the
emission region derived in the \citet{Eichler:1993} CWB model. For a
stellar separation of 26.45 AU at phase 0.737, and $\eta=0.22$, $\pi
r_O=26.6$AU.  The size of the emission region ($l$) and its projection
on the sky ($l_\perp$) is simply related by $l=l_\perp\sec\psi$, where
$\psi$ is defined above. At phase 0.737, the size observed by the VLBA
is 6~mas, which at 1.85 kpc corresponds to 11.1 AU. Since
$\psi=43\deg$ at this phase, $l=15.2$~AU, considerably smaller than
expected.  To attain the predicted size would require the observed
size to be 10.5~mas. As discussed earlier, the VLBA does resolve out
$\sim40$\% of the flux from the WCR in WR\thinspace140, so the
observed size is expected to be smaller than that predicted. However,
the MERLIN observation imposes an upper limit of 13 mas, consistent
with the predicted size.

\section{Astrometry of the WCR}
\begin{figure}
\vspace{11.4cm}
\includegraphics{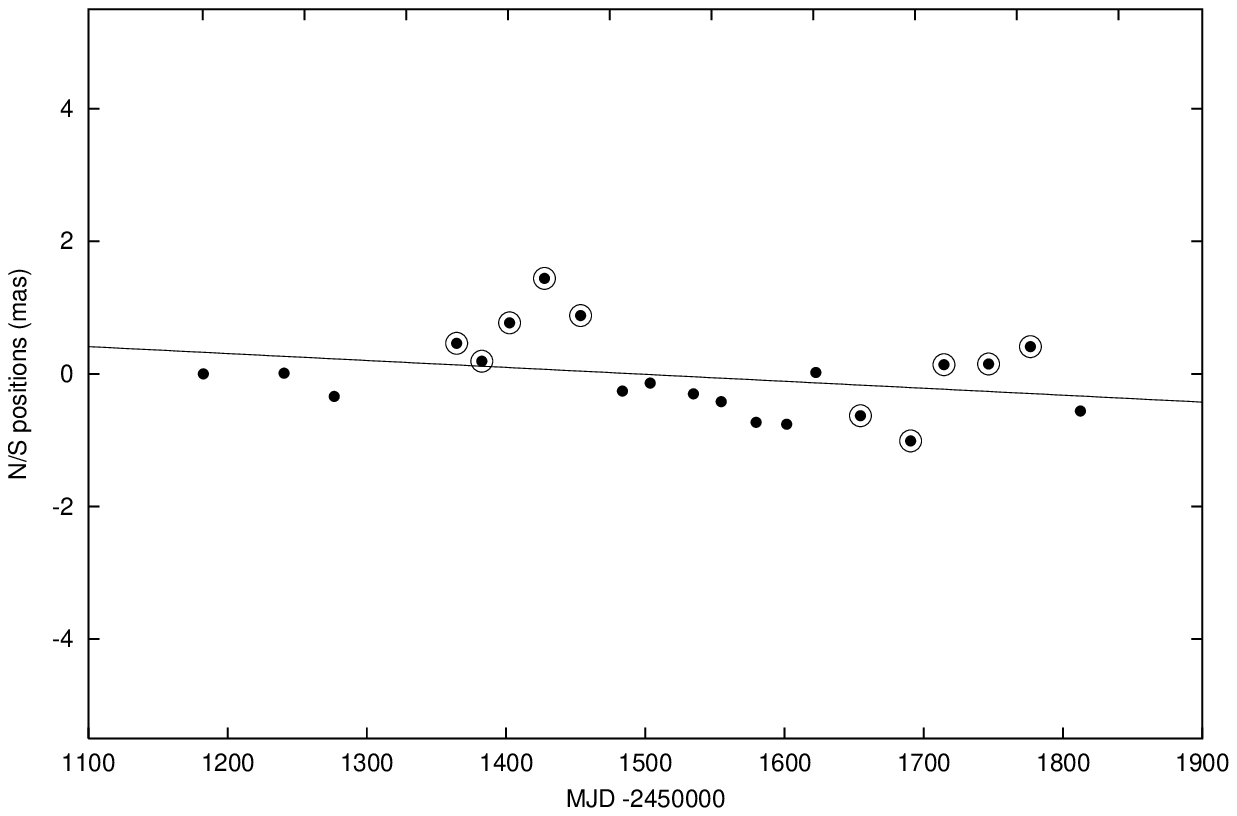}
\includegraphics{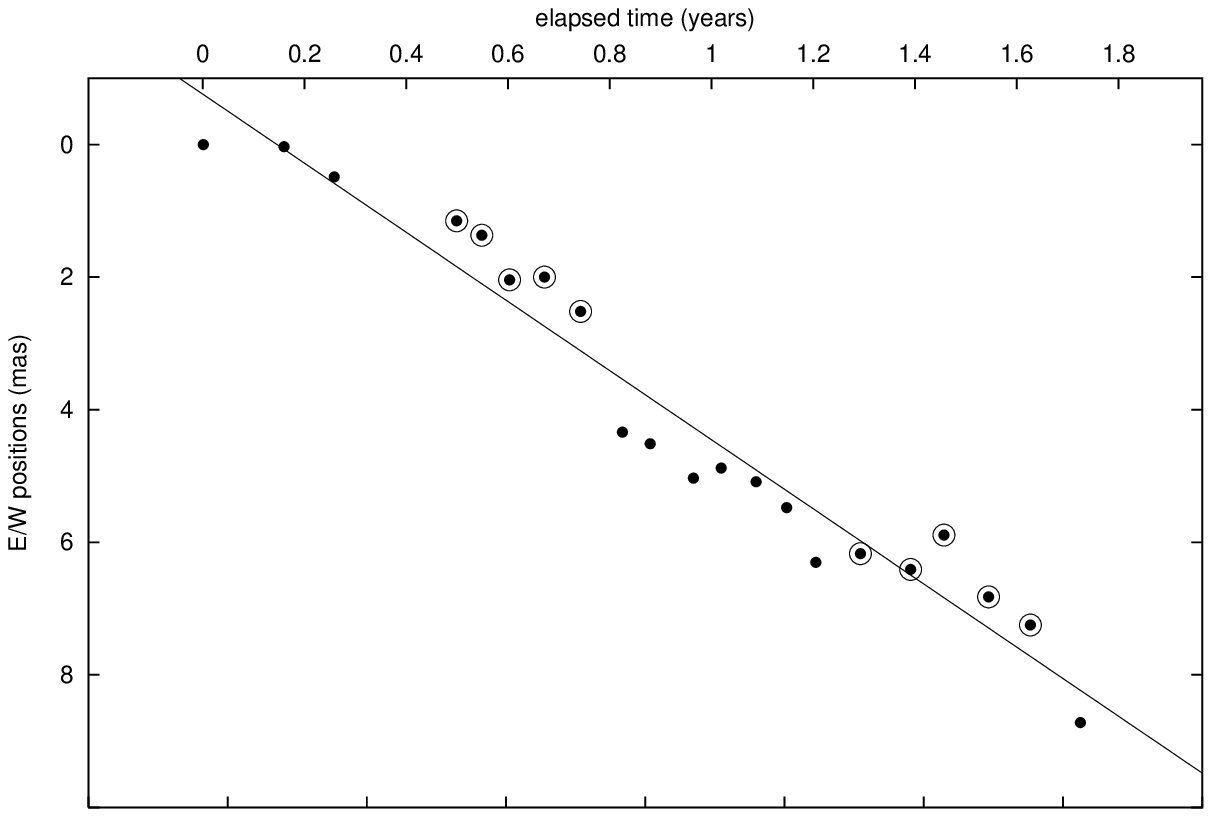}
\caption[]{The positions of the WCR at 8.4~GHz as a function of time
for motion east-to-west (top) and north-to-south (bottom). Both plots
have the same scales for easy comparison. Circled points represent
observations made between 1800-0600 MST.  Solid lines are regression
fits to all the data points, giving $\mu_\alpha=-5.20\pm0.27$~mas~yr$^{-1}$
and $\mu_\delta=-0.38\pm0.27$~mas~yr$^{-1}$. \label{fig:pm}}
\end{figure}

\subsection{Positions and proper motion of the WCR}
The use of phase-referencing permits accurate astrometry of the radio
emission of WR\thinspace140. The VLBA positions for the WCR are given
in Table~\ref{tab:vlba_obs}.  The offsets of the WCR in the 8.4~GHz
observations from the position at the first observation epoch as a
function of time are shown in Fig.~\ref{fig:pm}.  A largely westward
proper motion in the WCR is evident in the 8.4-GHz data.  A linear
regression fit to the 8.4-GHz positions gives a proper motion of
$\mu_\alpha=-5.20\pm0.27$~mas~yr$^{-1}$ and
$\mu_\delta=-0.38\pm0.27$~mas~yr$^{-1}$. In an attempt to assess if the
ionosphere has any clear impact on the 8.4-GHz astrometry,
observations between 1800 and 0600 MST (Socorro) are identified. The
east-west positions derived from night observations appear to be
eastward of the daytime positions. Similarly, the north-south
positions measured during the night are typically northward of day
time positions. We have no clear explanation for this apparent offset
in day versus night observations. However, repeating the
regression fits using only observations obtained during the night
gives a proper motion $\mu_\alpha=-5.44\pm0.25$~mas~yr$^{-1}$ and
$\mu_\delta=-0.84\pm0.46$~mas~yr$^{-1}$. The east-west proper motion is
closely the same as that derived from all data points, though the
derived north-south motion is larger, albeit with a large uncertainty.

Hipparcos observed a proper motion in WR\thinspace140 of
$\mu_\alpha=-5.36\pm0.58$~mas~yr$^{-1}$ and
$\mu_\delta=-2.37\pm0.49$~mas~yr$^{-1}$ between JD 2447870 and 2449046
\citep{Perryman:1997}. This date range spans orbital phase 0.26 to
0.67. Using the proper motion deduced for the WCR together with the
orbit defined here and $\eta=0.22$, the deduced proper motions of the
O and WR stars are $\mu_\alpha=-5.21$ and $-5.91$~mas~yr$^{-1}$ and
$\mu_\delta=-0.51$ and $-1.54$~mas~yr$^{-1}$ respectively.  The
deduced east-west motion of both the WR and O stars are consistent
with the Hipparcos proper motion. However, it would appear the deduced
north-south motion of the WR is closer than that of the O star to the
Hipparcos observation. This is at odds with the expectation that the
Hipparcos observations reflect the proper motion of the O star, since
the O star is expected to be the dominant emitter at visible
wavelengths (closely Johnson B and V) where the Hipparcos observations
were made.  However, given the large uncertainty in the Hipparcos
north-south proper motion, arguably both deduced values are consistent
with the Hipparcos observation within the measurement uncertainties.
\begin{figure}
\includegraphics[angle=0,scale=0.7]{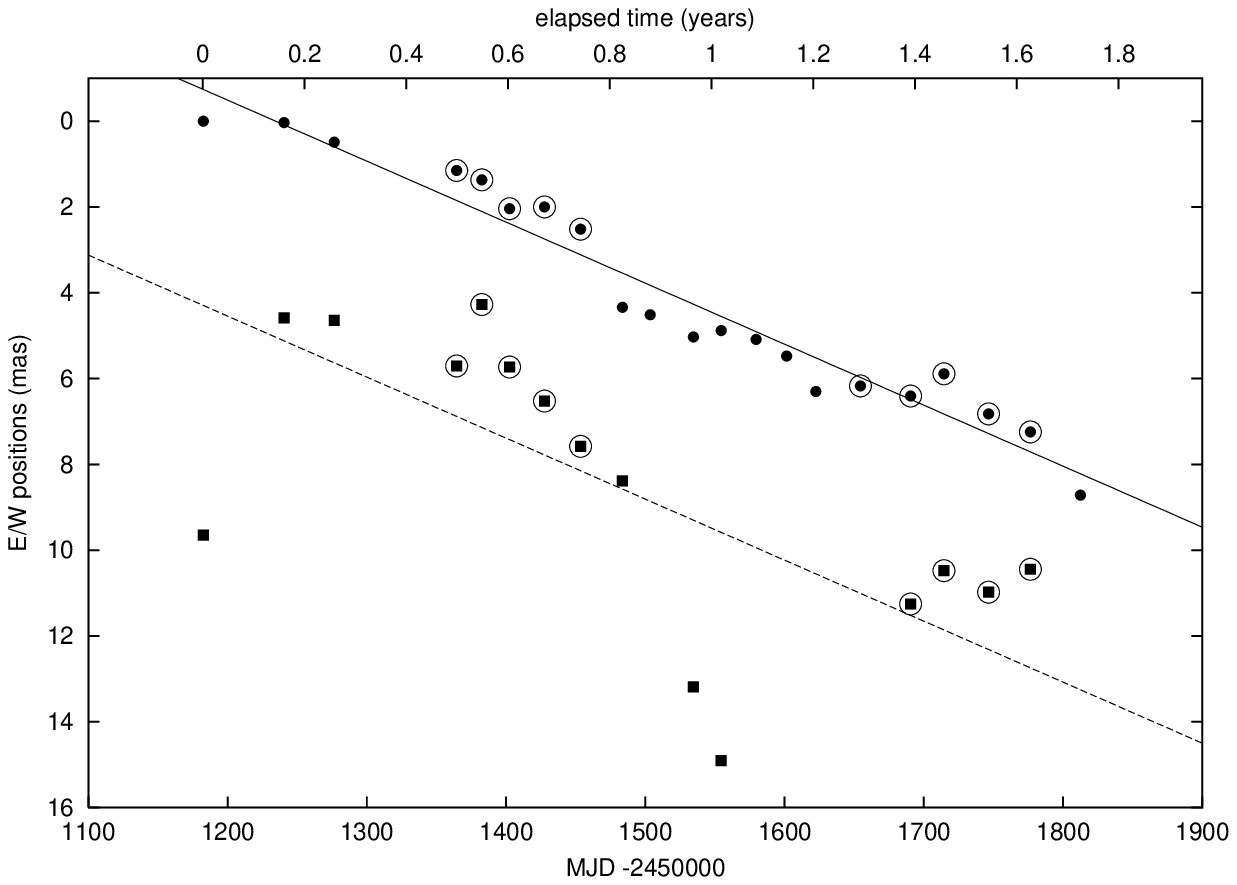}
\caption[]{The east-west proper motion of the WCR at 8.4~GHz (solid
circles) and 1.7~GHz (solid squares). An offset between the 1.7 and
8.4-GHz positions is clear. Circled points represent observations made
between 1800-0600 MST.  Solid lines are regression fits to all the
data points at each band, assuming $\mu_\alpha=-5.20$~mas~yr$^{-1}$, the
proper motion derived at 8.4~GHz. The offset in the two lines is
$5\pm1$~mas. \label{fig:offset}}
\end{figure}

\subsection{Offset between 8.4 and 1.7-GHz emission?}
In spite of the large scatter in the 1.7-GHz positions, the 1.7-GHz
emission appears to be westward of that at 8.4~GHz. In
Fig.~\ref{fig:offset} the 1.7 and 8.4-GHz proper motion in the
east-west direction is shown. From the figure, the offset in right
ascension at the two bands is readily apparent. We also note that the
largest excursions of the 1.7-GHz positions occur during daytime
observations, lending credibility to our suggestion that ionospheric
effects dominate the uncertainty in the astrometry. Taking
$\mu_\alpha=-5.20$ as derived from all the 8.4-GHz positions, an
offset of $5\pm1$ mas is derived between the 1.7 and 8.4-GHz right
ascension values. If we only use nighttime data at both bands along
with $\mu_\alpha=-5.44$, the position offset between the two bands is
$4\pm1$ mas.  We conclude that despite the poor astrometry at 1.7~GHz,
a right ascension offset of 4--5~mas between the peak of the 1.7 and
8.4-GHz emission is observed.

\section{The radio spectra of WR\thinspace140}

\begin{figure}
\includegraphics[angle=0,scale=0.7]{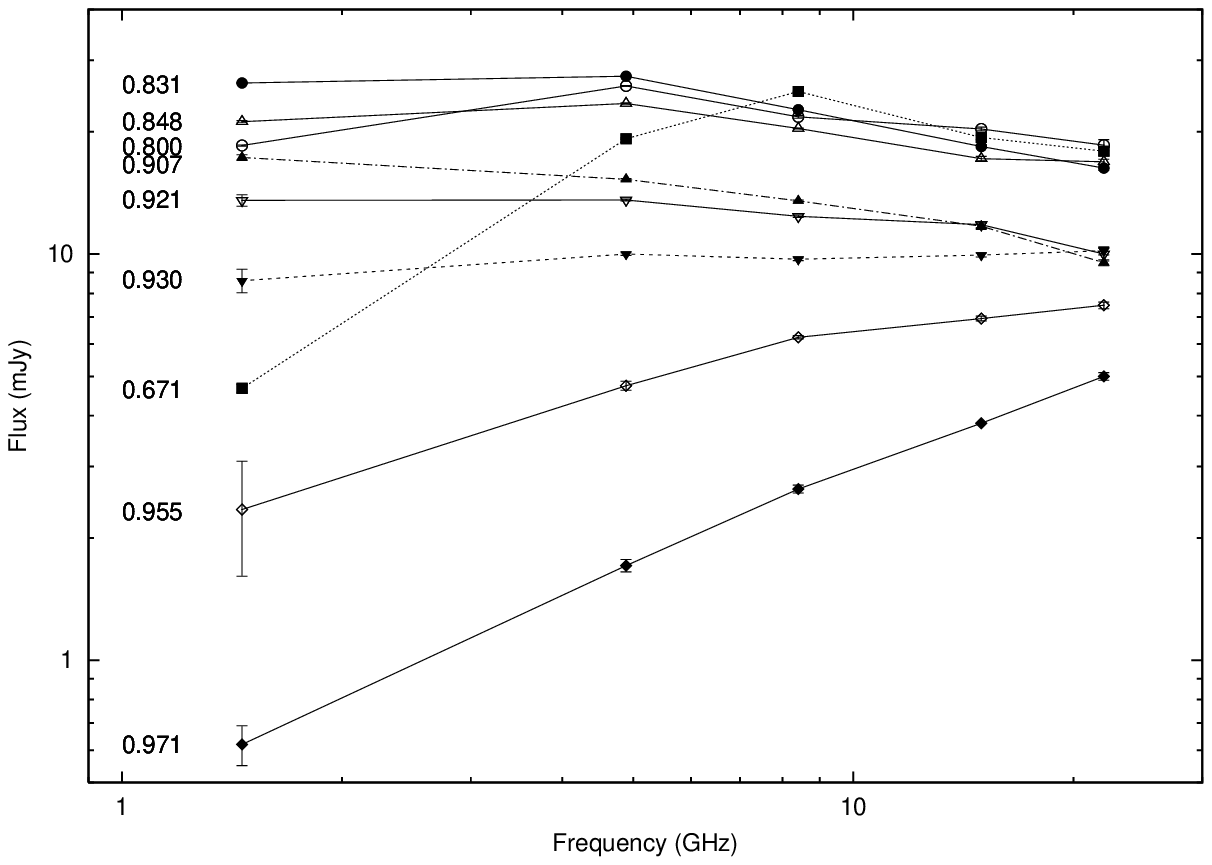}
\caption[]{The observed radio spectra of WR\thinspace140 at a
number of orbital phases as observed by the VLA. The orbital phase of 
each spectrum is shown on the left. Not all observed phases are
shown for clarity.\label{fig:new_spectra}}
\end{figure}

The new VLA data at five frequencies allow us to observe the radio
spectrum and its evolution better than previously possible, most
particularly at the higher frequencies. The radio spectra at a number
of orbital phases are shown in Fig.~\ref{fig:new_spectra} and exhibit
considerable evolution as the orbit progresses. At phase 0.67, the
spectrum shows a turnover at 8.4~GHz from a positive spectral
index\footnote{Flux $S_\nu\propto\nu^\alpha$, where $\alpha$ is the
spectral index} at low frequencies to a negative index at higher
frequencies. This is consistent with the data shown in
Fig.~\ref{fig:light_curve}, where a turnover between 15 and 5~GHz is
apparent around phase 0.65. Prior to this phase, the turnover
frequency is above 15~GHz.  As the orbital phase progresses after
phase 0.67, the turnover frequency moves to lower frequency, eventually to
below 1.5~GHz by phase 0.83, when the spectrum has a negative spectral
index across the observed frequencies.  The flux decreases at all
frequencies beyond this phase and the negative index remains until
phase 0.907, where the low frequency spectrum evolves rapidly.  By
phase 0.930, the spectrum is flat, and by 0.955 the turnover is
greater than 22~GHz, and the spectrum exhibits a positive index.

\begin{figure}
\includegraphics[angle=0,scale=0.7]{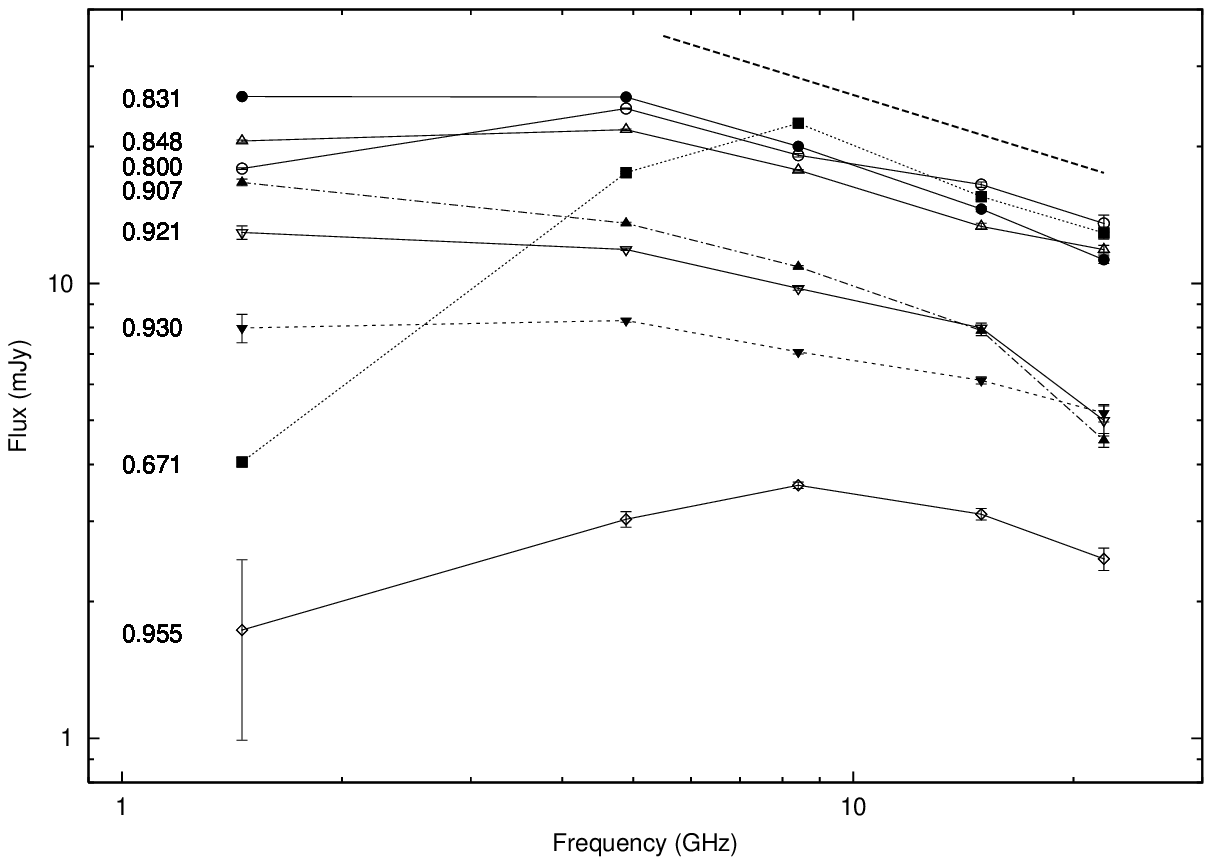}
\caption[]{Spectra of the synchrotron emission in WR\thinspace140 at
several orbital phases, determined by subtracting the thermal spectrum
at phase 0.974 (presumed to be solely from the stellar winds) from the
spectra at each phase.  The phase of each spectrum is shown on the
left.  The thick dashed line at the top indicates a slope of
$-0.5$.\label{fig:sync_spectra}}
\end{figure}

At phase 0.974, the spectrum is a power-law with a best-fit spectral
index of $0.72\pm0.03$, a value characteristic of the stellar winds in
WR+OB binary systems \citep{Williams:1996}. Comparison with the
observations from the previous orbit show that the radio flux at 5~GHz
is at its minimum value at this phase. The brightness temperature at
this phase is less than $10^5$~K (undetected by the VLBA), implying
thermal emission and we will assume the emission at this phase
represents the thermal emission from the stellar winds alone.
Certainly, the fluxes at phase 0.974 are consistent with the thermal
emission expected from the stellar winds of a WR+O binary with
T$\sim10^4$~K, mass-loss rates and terminal velocities as given in
Table~\ref{tab:stellar_parms}, with a filling factor $\sim0.15-0.3$,
similar to that derived earlier.  Assuming the thermal emission from
WR\thinspace140 is essentially constant throughout the orbit, the
synchrotron spectra at each observed phase can be determined by simply
subtracting the thermal flux at phase 0.974 from the total flux. No
15-GHz observation was made at phase 0.974 so a value of 3.8 mJy is
used, derived from the data of \citet{White:1995} and consistent with
the phase 0.974 spectrum shown in Fig.~\ref{fig:new_spectra}.  The
resulting synchrotron spectra are shown in
Fig.~\ref{fig:sync_spectra}.

The synchrotron spectra between phases 0.67 and 0.92 is optically thin
at several frequencies, with a spectral index that appears to be
closely constant with $\alpha=-0.5\pm0.1$, very similar to that
expected for first-order Fermi acceleration of electrons in strong,
non-relativistic shocks \citep[e.g. see][and references
therein]{Drury:1983, Jones:1991}.  The turnover frequency between the
optically thin and thick parts of the spectra appears to be related to
orbital phase. Between phase 0.67 and 0.85 the turnover is between 1.5
and 22~GHz, moving to lower frequencies as the orbit progresses. Prior
to 0.64, the turnover frequency is greater than 15~GHz (see
Fig.~\ref{fig:light_curve}) and after phase 0.85 it is less than
1.4~GHz. After phase 0.91, the spectra start to flatten as the
luminosity decreases. By phase 0.96 the turnover frequency is at
8.4~GHz.

The optically thick part of the synchrotron spectrum has been widely
attributed to free-free absorption in the stellar winds along the
line-of-sight to the WCR.  \citet{Williams:1990} found that if the
stellar wind envelope is assumed to be radially symmetric, the
free-free opacity to the apex of the WCR (i.e. near the stagnation
point of the two colliding winds) only exhibits large values and
variations close to periastron, when lines-of-sight to the apex of the
WCR have to traverse the inner regions of the WR star wind. Their
modelling of two-frequency data led them to suggest a more complex
geometry with the O-star envelope providing a lower density region
extending away from the WR star, very similar to the now generally
accepted geometry for the stellar wind envelopes in CWBs
\citep[c.f.][]{Eichler:1993, Dougherty:2003a}.  \citet{White:1995}
proposed an ad-hoc model to explain the radio peak near phase 0.7
using free-free opacity variations alone, where the WR wind is
equatorially enhanced and masks the O-star and the WCR for the bulk of
the orbit.  In this model, the rise to the radio peak at phase 0.7
indicates lower free-free opacity as the O-star passes through the
disk toward the observer.

Unfortunately, these models are too simple to explain the radio
observations of WR\thinspace140, as readily acknowledged by their
authors. The VLBA observations show the WCR as a distributed emission
region and the lines-of-sight to the WCR traverse different regions of
the stellar winds of the two stars. As a result, the emerging
emission will be a combination of both optically thick and thin
emission, since even though lines-of-sight to the apex may be
optically thick, a substantial amount of emission arises from
optically thin lines-of-sight to the downstream flow. Using the newly
derived orbit and assuming a half-opening angle for the WCR of
$63\deg$, we now know the lines-of-sight to the WCR traverse the
O-star wind between orbit phases 0.24 and 0.99
(Fig.~\ref{fig:losvsloc}) during which the most dramatic changes in
the radio emission are observed. Clearly, if stellar wind free-free
opacity is important it is the O-star wind opacity that is of most
concern, not that of the WR star.

Another shortcoming of previous models is the assumption the
synchrotron emission is optically thin at all observing frequencies,
and at all orbital phases. Though the high frequency data presented
here suggests an optically thin component at some orbital phases and
frequencies, it is necessary to account for the optically thick
component of the spectra observed during the bulk of the orbit. This
may, at least in part, to be due to mechanisms intrinsic to the
WCR. Certainly the models of both \citet{Williams:1990} and
\citet{White:1995} suggest a synchrotron source that is varying
dramatically throughout the orbit, variations that are modulated by
the orbital motion.

Before exploring the potential impact of a few mechanisms, it is
worthwhile estimating the equipartition magnetic field ($B_{eq}$) at
the apex of the WCR. The synchrotron luminosity ($L$) at phase 0.85
based on the spectrum in Fig.~\ref{fig:sync_spectra} is
$1.4\times10^{30}$ erg~s$^{-1}$, and for a source size ($l$) of 7 mas
at 1.85~kpc, $B_{eq}\sim 50$~mG at the apex of the WCR. Since
$B_{eq}\propto L^{2/7} l^{-6/7}$ \citep{Pacholczyk:1970}, our estimate
is a lower limit since the luminosity is a lower limit (the luminosity
will increase when observations at frequencies greater than 22~GHz are
incorporated). By phase 0.93, the source size (3~mas) and luminosity
($5.7\times10^{29}$~erg~s$^{-1}$) give $B_{eq}\sim75$~mG.
Encouragingly, both these estimates are the same order of magnitude as
those determined by \citet{Dougherty:2003b}.  It is interesting to
note that these magnetic field values are closely proportional to
$1/D$, as expected in a dipole field.

Inverse Compton (IC) cooling has been suggested as important in
WR\thinspace140, with the close proximity of the O star to the WCR
\citep{Dougherty:2003a}.  The IC cooling time at the apex of the WCR
is given by
$$t_{IC}=2.378\times10^{22} {r_O^2\over{L_O}}
\big({B\over{\nu}}\big)^{1/2},$$ where $r_O$ is the distance from the
O star to the WCR, $L_O$ the bolometric luminosity of the O star, and
$B$ is the magnetic field. Using the parameters in
Table~\ref{tab:stellar_parms}, at phase 0.85 the IC cooling time is
$\sim16$ hours at 22~GHz. This is short compared to the adiabatic
cooling time of $\sim90$ hours. Even at apastron, IC cooling dominates
($t_{ic}\sim170$~hours) over adiabatic cooling ($\sim430$~hours).
Since IC cooling $\propto r^{3/2}$ compared to adiabatic cooling
($\propto r$), at some point in the downstream flow adiabatic cooling
will dominate. Synchrotron cooling is not important in
WR\thinspace140, with timescales of $> 200$~days.  The signature of IC
cooling is a spectrum that is reduced at all frequencies, but most
severely at high frequencies, and is {\em not} characterised by a
high-frequency knee \citetext{Pittard et al. 2005, in
preparation}. This process may be the underlying cause of the
reduction in the luminosity as the orbit of WR\thinspace140 progresses
beyond 0.82, given its domination over the other cooling mechanisms.

The presence of thermal ions in the post-shock flow can result in a
departure of the refractive index of the plasma from unity, giving
rise to the Razin effect \citep[see e.g.][and references
therein]{Hornby:1966}. The frequency at which such a departure from
unity is significant is given by $\nu_R\approx 20 (n_e/B)$, where
$n_e$ and $B$ are the number density of electrons and the magnetic
field strength respectively. At phase 0.85 and for $\eta=0.22$, the
distance from the WR star to the WCR is 14.5 AU.  A mass-loss rate of
$4.2\times10^{-5} {\rm M}_\odot$~yr$^{-1}$ implies a post-shock
density in the WCR of $6.2\times10^6$~cm$^{-3}$. Taking $B$ to be
$50$~mG, $\nu_R\sim2.4$~GHz, within our observing range.  

For a field of 50~mG, a source of radius 3~mas and a flux of 15~mJy at
22~GHz in the optically thin part of the spectrum, synchrotron
self-absorption will produce a turnover ($\nu_s$) near 0.3~GHz,
outside the range of our observations. Since $\nu_s\propto B^{1/5}
R^{-4/5}$, which is very weakly dependent on the magnetic field, the
source would have to be $20-30\times$ smaller for self-absorption to
cause a turnover within the observed frequency range.

In highly eccentric systems such as WR\thinspace140, the relative
importance of the various emission, absorption and plasma processes
changes with stellar separation through variations in electron and
ion density ($\propto D^{-2}$), magnetic field strength ($\propto
D^{-1}$), and radiation field density ($\propto D^{-2}$). Given these
dependences, the asymmetry of the radio emission in WR\thinspace140
about periastron/apastron remains a puzzle - clearly it cannot be
explained by any one of these processes alone. The evolution of the
spectrum is most likely a result of the combined effects and relative
changes in these mechanisms as the orbit progresses. The full impact
of these effects on the observed emission from WR\thinspace140 awaits
more sophisticated models than used in previous work.

New radiative transfer models, based on a fully consistent
hydrodynamic treatment of the WCR, have started to explore the impact
of a number of processes on the radio emission from CWBs, including
free-free opacity in the stellar winds, synchrotron self-absorption,
Coulombic cooling through interactions with post-shock ions, plasma
effects such as the Razin effect, and Inverse Compton cooling by the
intense ultra-violet radiation field of the nearby massive
stars. These models have been very successful in explaining the radio
emission from very wide CWBs such as WR\thinspace 147
\citep{Dougherty:2003a}, and are now maturing to the point where they
will provide more insight to the mechanisms acting in systems like
WR\thinspace140 \citetext{Pittard et al. 2005, in preparation}. The
observations presented here represent the key constraints for these
new models.

\acknowledgments The authors thank the staff of the VLBA, the VLA and
MERLIN for their support.  The National Radio Astronomy Observatory is
a facility of the National Science Foundation operated under
cooperative agreement by Associated Universities, Inc. MERLIN is a
National Facility operated by the University of Manchester at Jodrell
Bank Observatory on behalf of PPARC. AJB gratefully acknowledges
support from NSF grant AST 0116558.  We'd also like to thank Julian
Pittard and Perry Williams for useful discussions and critical reading
of the manuscript, and Anita Richards for supplying the calibrated
MERLIN archive data.

\end{document}